\documentclass[a4paper,11pt]{article}
\pdfoutput=1
\usepackage{jheppub}
\usepackage{amsthm}
\usepackage{amsmath}
\usepackage{float}
\usepackage{graphicx}
\usepackage{subcaption}
\usepackage{slashed}
\usepackage{amssymb}
\usepackage{tensor}

\usepackage{physics}

\def\N{\mathcal{N}}

\begin{document}
\title{Thermal Emission of  Gravitational Waves from Weak to Strong Coupling}

\author{Luc\'ia Castells-Tiestos}
\author{and Jorge Casalderrey-Solana}
\affiliation{Departament de F\'\i sica Qu\`antica i Astrof\'\i sica \& Institut de Ci\`encies del Cosmos (ICC), 
Universitat de Barcelona, Mart\'{\i}  i Franqu\`es 1, 08028 Barcelona, Spain}

\emailAdd{lucia.castells@icc.ub.edu}
\emailAdd{jorge.casalderrey@ub.edu}

\abstract{We study the production of gravitational waves by a thermalized plasma of $\N $=4 Supersymmetric Yang Mills matter. We focus on the large number of colors limit, $N_c\rightarrow \infty$, and compute the spectrum of gravitational waves both for infinitely large and infinitesimally small values of the 't Hooft coupling constant $\lambda$. In the $\lambda\rightarrow \infty$ limit we employ the gauge/gravity duality to compute the emission rate via the analysis of Energy-Momentum tensor thermal correlators. In the $\lambda \rightarrow 0$ limit we employ state-of-the-art perturbative analyses to calculate the complete leading order emission rate. By comparing these extreme limits, we bracket the magnitude of the spectrum induced by this source of gravitational waves. Embedding our results in a cosmological evolution model, we find qualitative  and quantitative similarities between the strong coupling spectrum and the extrapolation of the perturbative results up to an intermediate value of the coupling, after an appropriate re-scaling of the effective number of degrees of freedom. We comment on how our results can help better understand the contribution of thermalized matter to the stochastic spectrum of gravitational waves.}
\maketitle

\section{Introduction}
The detection of gravitational waves (GW) by the LIGO and VIRGO collaboration~\cite{LIGOScientific:2016aoc} has initiated a new era in the exploration of the Universe. The measurements performed by those experiments have had a tremendous impact on our knowledge of compact objects in the Universe and provided tests of general relativity \cite{LIGOScientific:2020tif}. In addition, gravitational waves at different frequencies carry information of many different cosmological processes  at very different energy scales. These probes not only can help us better understand the early stages of the Universe, but can also provide new information on particle physics beyond the Standard Model (SM)~\cite{Barack:2018yly}.

Going beyond our current capabilities for detection, the high-frequency region $f \gtrsim 30$~kHz has recently attracted a lot of interest (see for example~\cite{Aggarwal:2020olq}). Gravitational waves at those frequencies are radiated by several different sources active both in the early and the late Universe (see~\cite{Caprini:2018mtu} for a review). While accessing this region requires the development of new technologies, the absence of known astrophysical sources at those frequency bands provides a strong motivation to study gravitational wave production in this challenging part of the spectrum. And one of the unavoidable sources that must be present in this frequency region is thermal emission of gravitational waves.

The fact that fluctuations in a thermalized system lead to the emission of gravitational waves is a well-known phenomenon that can be found in textbooks (see for example~\cite{weinberg1972gravitation}). From a quasi-particle point of view, this emission may be understood as the result of the collision of the plasma constituents. In analogy with bremsstrahlung photon radiation, those collisions lead to a gravitational wave spectrum with a characteristic scale given by the temperature $T$ of the plasma. Note that, as long as $T$ is small compared to the Planck mass, $T\ll M_\text{Pl}$, the emitted gravitational waves interact weakly with matter and therefore do not re-scatter with the plasma. Accordingly, the emitted GW radiation does not equilibrate at those temperatures. As a consequence, the spectrum of GW is not thermal, but contains information about the thermalized medium that produced those waves. This source of GW has been recently re-analysed in~\cite{ghiglieri2015gravitational,ghiglieri2020gravitational} by employing state-of-the-art perturbative analysis to compute the complete leading-order GW spectrum within SM. Building on those works, the authors of Ref.~\cite{ringwald2021gravitational} performed a survey of theories Beyond the Standard Model (BSM) and argued that this thermal emission may be used to constrain the maximum temperature and the effective number of degrees of freedom of the early Universe. For the SM as well as for the extensions studied in~\cite{ringwald2021gravitational}, this spectrum peaks in the vicinity of $f\sim 100$ GHz and is negligible at the much lower frequencies explored by current or already planed GW experiments.

A natural question to ask is how the features of this spectrum change by increasing orders in perturbation theory. However, going beyond leading-order in the calculation of this rate is a formidable task. As it is well known, the perturbative expansion in thermal field theory exhibits many complications which require performing different types of resummations (see~\cite{Ghiglieri:2020dpq} for a recent review). Nevertheless, the next-to-leading order analysis for a similar process (the thermal emission rate of photons by a QCD plasma) has been performed in~\cite{Ghiglieri:2013gia}. The analysis of that photon rate indicates that even for large values of the strong coupling constant, the main qualitative and quantitative features of the leading-order calculation remain the same at next-to-leading order. It would be interesting to complete the analysis for this channel.

A complementary way to assess the effect that matter interactions have on the production of GW is to study how much the spectrum changes when varying the magnitude of the coupling constant. A particularly interesting regime is the large coupling limit, when interactions among the plasma constituents make them lose their individual identity and the thermal excitations of the plasma cannot be described by quasi-particles. To perform such analysis we need to compute the emission rate by other means than perturbation theory. While we do not expect to find such regime within the SM, some of its extensions, for example composite models (see~\cite{Cacciapaglia:2020kgq} for a review) may possess a strongly-coupled phase.

While the strong coupling limit is challenging in general, there is an interacting gauge theory, $\N=4$ supersymmetric Yang Mills (SYM), for which we can access this regime. Since this theory is conformal, the 't Hooft coupling $\lambda\equiv g^2 N_c$ does not run and becomes a parameter of the theory which we can choose. In the large $\lambda$ limit, the so-called gauge/gravity duality~\cite{Maldacena:1997re} provides us with an alternative tool which allows us to address dynamical questions in the large number of colors $N_c\rightarrow\infty $, large 't Hooft coupling $\lambda \rightarrow \infty $ limits.

In this paper we use the gauge/gravity duality to perform the first calculation of the spectrum of thermally produced gravitational waves at strong coupling. We will contrast this strong coupling calculation with a complete leading-order analysis of the same quantity in perturbative $\N=4$ SYM. For this perturbative computation we will exploit the perturbative results obtained for a general $SU(N)$ gauge theory coupled to an arbitrary number of bosons and fermions presented in Refs.~\cite{ghiglieri2020gravitational,ringwald2021gravitational}.

This paper is organised as follows. In Section~\ref{gravitational-waves}, we review the weak-field approach to gravitational waves and the relation between the emission rate and a particular contraction of the thermal Energy-Momentum tensor correlators. In Section~\ref{TERSC}, we use the gauge/gravity duality to compute this correlator and determine the thermal emission rate of gravitational waves at strong coupling. In Section~\ref{comparison-weakly-coupled-source}, we perform the perturbative computation and compare it with the strong coupling analog for different values of the coupling constant. We later embed these two computations into a cosmological model in Section~\ref{SYM-in-cosmology} to compare the spectra after expansion. Finally, in Section~\ref{disc}, we summarize our main findings and we put our results into the general context of characterising this particular component of the stochastic GW background.

\section{Gravitational waves from a thermal source}\label{gravitational-waves}
In this section, we briefly review the existing literature on the emission of gravitational waves from a thermal source. The main physics of this process may be found in textbooks (see for example~\cite{weinberg1972gravitation}). To make a direct connection with the strong coupling calculation of Section~\ref{TERSC}, we will express the production rate in terms of a particular thermal correlator of the energy-momentum tensor. This discussion is based on~\cite{ghiglieri2015gravitational}, which we follow closely.

As it is well known in general relativity, gravitational waves at far-away distances from the emission source may be described under the weak-field approximation by small perturbations of flat spacetime $\eta_{\mu\nu}$, such that
\begin{equation}
    g_{\mu\nu}=\eta_{\mu\nu}+h_{\mu\nu}\;.
\end{equation}
The gauge freedom enjoyed by metric fluctuations can be removed by sticking to some particular gauge. In the so-called harmonic gauge, $h_{\mu\nu}$ satisfy the condition
\begin{equation}
    \frac{\partial}{\partial x^\mu}\tensor{h}{^\mu_\nu}=\frac{1}{2}\frac{\partial}{\partial x^\nu}\tensor{h}{^\mu_\mu}\;.
\end{equation}
Also, physical excitations that propagate far away from the source can be expressed in terms of the transverse traceless (TT) components of the fluctuating fields, $h_{\mu\nu}^\text{TT}$:
\begin{equation}
    k^\mu h^\text{TT}_{\mu\nu}=0\;,\phantom{abc}h^{\text{TT}\mu}_\mu=0\;,
\end{equation}
with $k^\mu=(\omega,\mathbf k)$ the four-momentum of the wave. Under these gauge conditions, the spatial Einstein equations become
\begin{equation}
    \Box h_{i j}^\text{TT}=-16\pi GT_{ i j}^\text{TT}\;,
    \label{eq:EinsteinEq}
\end{equation}
where $\Box\equiv -\partial_t^2+\nabla^2$ and $T_{ij}^\text{TT}$ are the spatial components of the transverse traceless part of the energy-momentum tensor. After a Fourier transform, $T_{\mu\nu}^\text{TT}$ may be obtained as
\begin{equation}
T_{ij}^\text{TT}(k^\mu)=\Lambda_{ijmn}T_{mn}(k^\mu)\;,
\label{eq:TTT_def}
\end{equation}
where we have introduced a projection operator onto the transverse traceless modes
\begin{equation}
    \Lambda_{ijmn}\equiv\frac{1}{2}\left(P_{im}P_{jn}+P_{in}P_{jm}-P_{ij}P_{mn}\right)\;,
\end{equation}
with $P_{ij}$ the spatial components of the transverse projector 
\begin{equation}
    P_{\mu\nu}=\eta_{\mu\nu}-\frac{k_\mu k_\nu}{k^2}\;.
    \label{eq:P-conserved-vectors}
\end{equation}

A solution to Eq.~\eqref{eq:EinsteinEq} can be easily found in momentum space as
\begin{equation}
h_{i j}^\text{TT} (\omega, \mathbf{k}) = 16\pi G \, \frac{T_{ij}^\text{TT}(\omega, \mathbf{k})}{(\omega+i\epsilon)^2-\mathbf{k}^2} \,,
\end{equation}
where the $i \epsilon$-prescription selects the retarded solution. Fourier transforming the frequency, this solution may be written as
\begin{equation}
h_{ij}^\text{TT}(\mathbf{k},t)=16\pi G\int_{-\infty}^{t}dt'\;\frac{\sin \left(\omega(t-t')\right)}{\omega}\;T_{ij}^\text{TT}(\mathbf{k},t')\;.
\label{eq:hijTT}
\end{equation}

Since the $h_{ij}^\text{TT}$ components possess all the physical information of the gravitational field excitation, the classical energy carried away from the source by gravitational waves may be expressed in terms of $h_{ij}^\text{TT}$ as
\begin{equation}
    E_{\text{GW}}=\frac{1}{32\pi G}\int \frac{d^3k}{(2\pi)^3}\left[\dot h_{ij}^\text{TT}(-\mathbf{k},t) \, \dot h_{ij}^\text{TT}(\mathbf{k},t)\right]\;.
    \label{eq:classical-energy}
\end{equation}

Substituting the time-derivative of Eq.~\eqref{eq:hijTT} into this expression yields
\begin{align}
     E_{\text{GW}}&=8\pi G\int \frac{d^3k}{(2\pi)^3}\int_{-\infty}^t dt'\int_{-\infty}^t dt''\;\cos(\omega(t-t'))\cos(\omega(t-t''))T_{ij}^\text{TT}(-\mathbf{k},t')T_{ij}^\text{TT}(\mathbf{k},t'')\,. 
\end{align}
As it is standard in the classical computation of the radiation, the energy carried by gravitational waves is obtained after averaging the instantaneous energy $E_\text{GW}$ over an observation period that is long as compared to the frequency of the wave:
\begin{equation}
    \bar E_{\text{GW}}=\frac{1}{\tau}\int_{t-\frac{\tau}{2}}^{t+\frac{\tau}{2}}dt'\;E_{\text{GW}}\,.
    \label{eq:classical-energy2}
\end{equation}
After this observation-time average, the radiated energy is given by
\begin{align}
    \bar E_{\text{GW}}
    &=4\pi G\int \frac{d^3k}{(2\pi)^3}\int_{-\infty}^t dt'\int_{-\infty}^t dt''\;\cos(\omega(t'-t'')) T_{ij}^\text{TT}(\mathbf{k},t')T_{ij}^\text{TT}(-\mathbf{k},t'')\;. \label{eq:ETTgen}
\end{align}
This expression is valid for any source; in particular, it is valid for any of the individual configurations of the thermal ensemble. To determine the energy emitted by the thermal system, we average over those configurations with the thermal weight. Assuming that the theory is space and time translation invariant, we express the emitted energy in terms of the correlator
\begin{align}
\mathcal C(\mathbf{k},t'-t'') &\equiv \big\langle T_{ij}^\text{TT}(\mathbf{k},t')T_{ij}^\text{TT}(-\mathbf{k},t'')\big\rangle
\notag\\
    &=V\int d^3 x\;\big\langle T_{ij}^\text{TT}(\mathbf{x},t'-t'')T_{ij}^\text{TT}(0,0)\big\rangle\;e^{-i\mathbf{k\cdot x}}\,,
    \label{eq:Cdef}
\end{align}
where the brackets represent the thermal average and $V$ is the volume of the system. After averaging over all configurations and taking a time-derivative in Eq.~(\ref{eq:ETTgen}), the mean radiated energy per unit time becomes
\begin{align}
    \frac{d\bar E_{\text{GW}}}{dt}&=4\pi G\int \frac{d^3k}{(2\pi)^3}\int_{-\infty}^\infty d\tau\;e^{i\omega\tau}\;\frac{\mathcal C(\mathbf{k},\tau)+\mathcal C(\mathbf{k},-\tau)}{2}\;.\label{eq:power}
\end{align}
Introducing the definition of $\mathcal C(\mathbf{k},\tau)$ given by Eq.~\eqref{eq:Cdef} and using the time translation invariance of the average ensemble, we express the energy density $\rho_{\text{GW}}$ of emitted gravitational waves per unit time as
\begin{equation}
    \frac{d\rho_{\text{GW}}}{dtd^3\mathbf{k}}=\frac{4\pi G}{(2\pi)^3}\int d^4x\;e^{i(\omega t-\mathbf{k\cdot x})}\bigg\langle\frac{1}{2}\{T_{ij}^\text{TT}(\mathbf{x},t),T_{ij}^\text{TT}(0,0)\}\bigg\rangle\;,
    \label{eq:production-rate-classical}
\end{equation}
where $\{A,B\}\equiv AB+BA$. As anticipated, this classical analysis relates the rate of emission of gravitational waves to a correlation function in medium. However, the symmetrized appearance of the correlator is an artifact of the classical approximation, since in this limit the order of operators does not matter. As argued in~\cite{ghiglieri2015gravitational}, and in analogy with the thermal emission rate of  (weakly-coupled) photons, the full calculation is expressed in terms of a particular Wightman function. Recalling the definition of $T_{ i j}^\text{TT}$ given in Eq.~(\ref{eq:TTT_def}), and since $\Lambda_{ijmn}$ is a projector, the energy density of gravitational waves per unit time is given by
\begin{equation}
    \frac{d\rho_\text{GW}}{dtd^3\mathbf k}=\frac{4\pi G}{(2\pi)^3}\;\Lambda_{ijmn}\int d^4x\;e^{i(\omega t-\mathbf{k\cdot x})}\left\langle T^{ij}(0,0)T^{mn}(\mathbf x,t)\right\rangle\;.
    \label{eq:production-rate}
\end{equation}

The expression for the emission rate that we have just derived may be apparently formal, but it has a convenient form for our purposes: it explicitly separates the weakly-interacting dynamics of the gravitational waves emitted by the thermal plasma from the self-interaction of the matter in the plasma. While the calculation has been performed in the weak-field approximation for the gravitational disturbance, we have made no assumption on the dynamics of the plasma, other than some general symmetries, such as spacetime translation invariance. In the next section we will evaluate the thermal correlator in different approximations to the plasma dynamics.

\section{Thermal emission rate at strong coupling}\label{TERSC}
In this section, we evaluate the thermal correlator appearing in Eq.~\eqref{eq:production-rate}, and therefore the emission rate, in a strongly-coupled theory. In particular, we will focus on interacting gauge theories with a large value of the coupling strength. This is a complicated limit to analyse theoretically, since there are not many computational methods available. Certainly, perturbative methods are not generally applicable in this limit. Similarly, lattice field theory computations, even for theories with a well understood lattice implementation, are also of  limited guidance, since the emission rate equation~\eqref{eq:production-rate} is controlled by a light-like correlator. For this reason, in this paper we focus on $\N=4$ SYM at strong 't Hooft coupling $\lambda\rightarrow\infty$ (SCSYM), since for this theory we have a framework with which to address the strongly-coupled regime: the gauge/gravity duality. In this section we use this framework to compute the emission rate.

\subsection{Holographic analysis of the thermal emission rate}\label{gauge-string-duality}
The gauge/gravity duality, also known as holography, is a relation between a quantum field theory and a theory of gravity in at least one extra dimension (see~\cite{Aharony:1999ti} for a review). The earliest example of this duality, the so-called AdS/CFT correspondence~\cite{Maldacena:1997re}, establishes a relation between $\N=4$ $SU(N_c)$ SYM in the large number of colors $N_c\rightarrow \infty$ and the large 't Hooft coupling $\lambda\rightarrow\infty$ limits, for $\lambda\equiv g^2_{\text{YM}}N_c$, and type IIB supergravity on an AdS$_5\times$S$^5$ background, whose metric is given by
\begin{equation}
    ds^2=\frac{r^2}{R^2}\left(-f(r)dt^2+d\mathbf x^2\right)+\frac{R^2}{r^2f(r)}dr^2+R^2d\Omega_5^2\;,
    \label{eq:AdS-CFT-metric}
\end{equation}
with $R$ the radius of AdS  and $d\Omega_5^2$ the metric of a unit five-sphere. For the vacuum of $\N=4$, the function $f(r)=1$. For a thermal system, the function $f(r)$ transforms into $f(r)=1-r_0^4/r^4$ and Eq.~\eqref{eq:AdS-CFT-metric} therefore accounts for the black brane metric, with $r_0$ representing the position of the event horizon. In the latter case, the temperature of the dual field theory is the Hawking temperature of the black brane, $T=r_0/\pi R^2$. Since the observables we describe in this paper are not charged under the $R$-current, we will not consider excitations of the  five-sphere and, therefore, it will suffice to restrict ourselves to the five-dimensional part of the background. Introducing a new radial coordinate $u=r_0^2/r^2$, the metric may be rewritten as
\begin{equation}
    ds^2_{\text{AdS}_5}\equiv g_{\mu\nu}^{(0)}dx^\mu dx^\nu=\frac{(\pi TR)^2}{u}\left(-f(u)dt^2+d\mathbf x^2\right)+\frac{R^2}{4u^2f(u)}du^2\;.
    \label{eq:SCSYM-metric}
\end{equation}

According to the dictionary of the gauge/gravity duality, bulk fields in the gravity theory couple to local operators in the dual quantum field theory. More specifically, for each possible operator $\mathcal O(x)$ and its corresponding point-dependent source $\phi(x)$, we can assign a dual bulk field $\Phi(x,u)$ whose value at the boundary identifies with $\phi(x)$~\cite{Aharony:1999ti,Witten:1998qj,Gubser:1998bc}.

One of the important implications of this relation between bulk fields and operators is the possibility of calculating the $n$-point function of local operators in the gauge theory in terms of a gravity description~\cite{Aharony:1999ti,Witten:1998qj,Gubser:1998bc}. For instance, the general strategy for computing the retarded two-point function of a local, gauge-invariant operator $\mathcal O$ in terms of the gravity prescription is as follows \cite{Son:2002sd}: identify the bulk field $\Phi$ dual to the operator $\mathcal O$; find the string action and the corresponding equation of motion satisfied by $\Phi$ and solve it subject to the in-falling boundary condition; express the solution $\Phi_c$ in the basis of two local solutions at the boundary; the connection coefficients of this linear combination determine the retarded Green's function for $\mathcal O$.

This statement connects directly with the result we have derived in Eq.~\eqref{eq:production-rate}: the production rate of gravitational waves from a thermal source depends on the two-point correlation function of the energy-momentum tensor. In this way, the gauge/gravity duality allows us to simplify the calculation of the energy production rate by a thermal ensemble of strongly-coupled matter by simply transforming it into a general relativity problem. We only need to apply the correspondence to the concrete case of the energy-momentum tensor two-point function.

According to the duality, the energy-momentum tensor is dual to the metric of the gravitational theory. Thermal energy-momentum tensor correlators are determined by considering small gravitational perturbations, $h_{\mu\nu}$, over the black brane metric. Given that we are using the same notation, and to avoid possible confusions, it is important to point out at this point that these metric perturbations are different from the gravitational wave fluctuating fields described in Section~\ref{gravitational-waves}. In the latter case, those are fluctuations over the four-dimensional metric on the field theory side and represent dynamical fields that interact weakly with the strongly-coupled matter; on the former, the metric fluctuations encode modifications of the 5-dimensional metric dual to a state in the field theory and, from the point of view of the field theory, may be viewed as a calculational tool.

Before we describe the dynamics of the metric fluctuations in AdS$_5$, let us classify the possible excitations that we may encounter. Without loss of generality, we shall assume these fluctuations to propagate along the $z$-axis and to be of the form
\begin{equation}
    h_{\mu\nu}(t,\mathbf{x},u)=h_{\mu\nu}(u)e^{-i\omega t+ikz}\;,
    \label{eq:fluctuating-fields}
\end{equation}
with momentum $\mathbf k=(0,0,k)$. With this assumption, metric perturbations can be classified according to their transformation properties under rotations in the $xy$-plane, whether they transform as vectors (shear channel), as tensors (tensor channel) or remain invariant (sound channel)~\cite{kovtun2005quasinormal}:
\begin{subequations}
\begin{align}
    &\text{Sound channel (spin 0)}:\phantom{ab}h_{tt},\;h_{tz},\;h_{zz},\;h,\;h_{uu},\;h_{tu},\;h_{zu}\label{eq:gauge-inv-3}\\
    &\text{Shear channel (spin 1)}:\phantom{ab}h_{t\alpha},\;h_{z\alpha},\;h_{u\alpha}\label{eq:gauge-inv-1}\\
    &\text{Tensor channel (spin 2)}:\phantom{ab}h_{\alpha\beta}-\delta_{\alpha\beta}h/2\;,\label{eq:gauge-inv-2}
\end{align}
\label{eq:gauge-invariant}%
\end{subequations}
with $\alpha,\beta=x,y$. These excitations are governed by Einstein equations to linear order in $h$. Since Einstein equations and the black brane state are both invariant under rotations, the equations obeyed by each of these three sets of fluctuation components decouple from each other, making it easy to evaluate them separately.

Each of the symmetry channels has its own contribution to the energy-momentum tensor correlation function. In this way, in a conformal quantum field theory, the retarded two-point function of the energy-momentum tensor in a thermal rotation-invariant state is given by a sum of three independent components~\cite{kovtun2005quasinormal}
\begin{align}
    G_{\mu\nu\alpha\beta}^R(k)&=-i\int d^4x\;e^{i(kt-\mathbf{kx})}\left\langle\theta(t)\left[T_{\mu\nu}(0,0),T_{\alpha\beta}(\mathbf x,t)\right]\right\rangle\notag\\
    &=S_{\mu\nu\alpha\beta}G_1(k)+Q_{\mu\nu\alpha\beta}G_2(k)+L_{\mu\nu\alpha\beta}G_3(k)\;,
\end{align}
with $S_{\mu\nu\alpha\beta},Q_{\mu\nu\alpha\beta},L_{\mu\nu\alpha\beta}$ appropriate orthogonal projectors providing three different Lorentz index structures and $G_1,G_2,G_3$ three scalar functions determining the shape of the correlator. Those  Lorentz structures may be constructed in terms of the transverse and longitudinal parts of the projector~\eqref{eq:P-conserved-vectors}, defined as
\begin{subequations}
\begin{align}
    &P_{00}^T=P_{0i}^T=0\;,\phantom{abc}P_{ij}^T=\delta_{ij}-\frac{k_i k_j}{\mathbf{k}^2}\;,\\
    &P_{\mu\nu}^L=P_{\mu\nu}-P_{\mu\nu}^T\;,\phantom{abc}P_{\mu\nu}=g_{\mu\nu}-\frac{k_\mu k_\nu}{k^2}\;.
\end{align}
\end{subequations}
The three relevant, mutually orthogonal combinations of $P_{\mu\nu}^T$ and $P_{\mu\nu}^L$ are the following:
\begin{subequations}
\begin{align}
    S_{\mu\nu\alpha\beta}&=\frac{1}{2}\left(P_{\mu\alpha}^T P_{\nu\beta}^L+P_{\mu\alpha}^L P_{\nu\beta}^T+P_{\mu\beta}^T P_{\nu\alpha}^L+P_{\mu\beta}^L P_{\nu\alpha}^T\right)\;,\label{eq:projector-S}\\
    Q_{\mu\nu\alpha\beta}&=\frac{1}{3}\left(2P_{\mu\nu}^L P_{\alpha\beta}^L+\frac{1}{2}P_{\mu\nu}^T P_{\alpha\beta}^T-(P_{\mu\nu}^T P_{\alpha\beta}^L+P_{\mu\nu}^L P_{\alpha\beta}^T)\right)\;,\label{eq:projector-Q}\\
    L_{\mu\nu\alpha\beta}&=H_{\mu\nu\alpha\beta}-S_{\mu\nu\alpha\beta}-Q_{\mu\nu\alpha\beta}\;,\label{eq:projector-L}
\end{align}
\end{subequations}
with $H_{\mu\nu\alpha\beta}$ a projector onto traceless symmetric tensors
\begin{equation}
    H_{\mu\nu\alpha\beta}=\frac{1}{2}\left(P_{\mu\alpha} P_{\nu\beta}+P_{\mu\beta} P_{\nu\alpha}\right)-\frac{1}{3}P_{\mu\nu} P_{\alpha\beta}\;.
\end{equation}
The first combination~\eqref{eq:projector-S} projects onto the vector modes (shear channel), whereas the second one~\eqref{eq:projector-Q} projects onto the scalar modes (sound channel) and the third one~\eqref{eq:projector-L} onto the tensor modes (tensor channel). The corresponding scalar functions $G_1,G_2,G_3$ therefore couple to the spin 1, spin 0 and spin 2 channels, respectively. Finding a solution to the second-order ODEs obeyed by the variables in each of these channels allows us to compute the scalar functions $G_i$ and thus also the retarded energy-momentum tensor correlation function.

However, one should notice that the energy production rate~\eqref{eq:production-rate} we are ultimately interested in does not depend on the retarded correlator, but rather on the spatial components of the Wightman function, which we shall denote as
\begin{equation}
    G_{\mu\nu\alpha\beta}^<(k)=\int d^4x\;e^{i(kt-\mathbf{kx})}\left\langle T_{\mu\nu}(0,0)T_{\alpha\beta}(\mathbf x,t)\right\rangle\;.
\end{equation}
This correlator may be determined by making use of the Kubo-Martin-Schwinger (KMS) relations~\cite{laine2016basics}, which allow us to write it in terms of the spectral function
\begin{equation}
    \rho_{\mu\nu\alpha\beta}(k)=\int d^4x\;e^{i(kt-\mathbf{kx})}\left\langle\left[T_{\mu\nu}(0,0),T_{\alpha\beta}(\mathbf x,t)\right]\right\rangle\;,
\end{equation}
which, in turn, is proportional to the imaginary part of the retarded correlator as
\begin{equation}
    \rho_{\mu\nu\alpha\beta}(k)=-2\Im G_{\mu\nu\alpha\beta}^R\;.
    \label{eq:retarded-spectral}
\end{equation}
Using the KMS relations, the Wightman function is related to the spectral function as
\begin{equation}
    G_{\mu\nu\alpha\beta}^<(k)=n_B\rho_{\mu\nu\alpha\beta}(k)\;,
    \label{eq:wightman-spectral}
\end{equation}
where $n_B=1/\left(\exp[k/T]-1\right)$.
 
Writing Eqs.~\eqref{eq:retarded-spectral} and~\eqref{eq:wightman-spectral} together yields the relation between both correlation functions:
\begin{equation}
    G_{\mu\nu\alpha\beta}^<(k)=-2n_B\Im G_{\mu\nu\alpha\beta}^R=-2n_B\Im\left(S_{\mu\nu\alpha\beta}G_1+Q_{\mu\nu\alpha\beta}G_2+L_{\mu\nu\alpha\beta}G_3\right)\;.
\end{equation}
Introducing this expression into Eq.~\eqref{eq:production-rate} and computing the projectors for $\mathbf k=(0,0,k)$, the energy production rate reduces to
\begin{equation}
    \frac{d\rho_\text{GW}}{dtd^3\mathbf k}=\frac{-16\pi Gn_B}{(2\pi)^3}\Im G_3\;,
    \label{eq:production-rate-G3}
\end{equation}
or, equivalently expressed per unit logarithmic wave number interval,
\begin{equation}
    \frac{d\rho_\text{GW}}{dtd\log k}=\frac{-16\pi G n_Bk^3}{(2\pi)^3}\Im G_3\;.
    \label{eq:log-production-rate-G3}
\end{equation}
Here we have used that $\Lambda_{ijmn}S^{ijmn}=\Lambda_{ijmn}Q^{ijmn}=0$ and $\Lambda_{ijmn}L^{ijmn}=2$, which implies that the energy-momentum tensor in a thermal gravitational-wave system only couples to fluctuating fields in the tensor channel~\eqref{eq:gauge-inv-2}. For this reason, to compute the GW production rate, it will suffice to restrict ourselves to the study of the tensor channel and keep only off-diagonal perturbations $h_{xy}$.

\subsection{Computation of the emission rate}\label{strongly-coupled-SYM-theory}
In this section, we study the dynamics of tensor fluctuations over the black brane solution~\eqref{eq:SCSYM-metric}. This type of fluctuations has been extensively studied in the past (see for example~\cite{kovtun2005quasinormal}). Here we will follow the general analysis of~\cite{Teaney:2006nc,Kovtun:2006pf} and use it to numerically determine the spectral function $\rho$, defined in Eq.~\eqref{eq:retarded-spectral}, for light-like momenta.

The dynamics of tensor fluctuations about the black brane metric are governed by the quadratic approximation to the gravitational action 
\begin{equation}
    S=\frac{1}{16\pi G_5}\left[\int d^5x\;\sqrt{-g}\left(\mathcal R+2\Lambda\right)+2\int d^4x\;\sqrt{-\partial g}K\right]\;,
    \label{eq:SCSYM-action}
\end{equation}
with $G_5$ the effective five-dimensional Newton constant, $\Lambda\equiv 6/R^2$ the cosmological constant of the AdS spacetime and $K$ the extrinsic curvature over the induced boundary metric $\partial g$. As a consequence of the duality, $G_5$ is related to the parameters of the field theory as
\begin{equation}
    \frac{G_5}{R^3}=\frac{\pi}{2N_c^2}\;,
    \label{eq:G5}
\end{equation}
where $N_c$ is the number of colors of the dual $SU(N_c)$ gauge theory.

Either by extremizing the gravitational action~\eqref{eq:SCSYM-action} or by linearizing the Einstein equations, the equations for fluctuations $h_{\mu\nu}$ of the form~\eqref{eq:fluctuating-fields} may be written as
\begin{equation}
    \mathcal R_{\mu\nu}^{(1)}=-\frac{4}{R^2}h_{\mu\nu}\;,
    \label{eq:linear-eom}
\end{equation}
with $\mathcal R_{\mu\nu}^{(1)}$ the Ricci tensor to linear order in the perturbation $h$. 

As it is well known, diffeomorphism invariance implies that $h$ possesses a gauge symmetry and, therefore, not all solutions to these equations correspond to physical excitations. For this reason, it is convenient to remove this gauge freedom by working with gauge-invariant variables constructed as linear combinations of the fields and their derivatives in the corresponding symmetry channel~\eqref{eq:gauge-invariant}. Since we are only interested in spin 2 variables, we will restrict ourselves to the gauge-invariant variable in the tensor channel~\cite{kovtun2005quasinormal}:
\begin{equation}
    Z(u)=h_{y}^{x}(u)\;.
\end{equation}
Introducing this expression into the linearized equation of motion~\eqref{eq:linear-eom} for $\mu=x,\;\nu=y$, yields the following second-order differential equation in terms of the new variable $Z$:
\begin{equation}
    Z''-\frac{1+u^2}{uf}Z'+\frac{\mathfrak w^2-\kappa^2f}{uf^2}Z=0\;, \label{eq:ODE-Z3}
\end{equation}
with the dimensionless parameters $\mathfrak w$ and $\kappa$ defined as $\mathfrak w=\omega/2\pi T$ and $\kappa=k/2\pi T$, with $\omega$ and $k$ the frequency and momentum of the excitation. Since gravitational waves are controlled by light-like correlators, in the rest of the analysis we will set $\mathfrak w=\kappa$.

The general strategy for computing the two-point correlation function follows the steps we have outlined in Section~\ref{gauge-string-duality}. Following that discussion, we search for a solution of this equation satisfying the incoming-wave boundary condition. However, given that it is not possible to find analytic solutions to the equation~\eqref{eq:ODE-Z3}, we will solve it through numerical methods by performing a series expansion near the boundary and the horizon.

The above ODE~\eqref{eq:ODE-Z3} has two singular points at $u=0,1$. Following~\cite{morse1954methods}, we can work out a Frobenius series solution about each of these singularities. Near the horizon $u=1$, the series solution takes the form
\begin{equation}
    Z(u)=\left(1-u^2\right)^{\pm i\kappa/2}F_H(u)\;,
    \label{eq:Z1}
\end{equation}
with $F_H(u)$ a power series about $u=1$. The minus sign corresponds to the incoming-wave solution, while the plus sign is associated to the outgoing-wave solution. Since we demand $Z$ to satisfy the incoming-wave boundary condition, we will reject the solution with a positive sign in the exponent. The expression for $F_H(u)$ can be inferred by demanding Eq.~\eqref{eq:Z1} to satisfy the equation of motion~\eqref{eq:ODE-Z3}. In this way, we may find a power series  solution near the horizon:
\begin{align}
    Z_1=\left(1-u^2\right)^{-i\kappa/2}\left[1-(1-u)\left(\frac{i\kappa^2}{4(i+\kappa)}\right)-(1-u)^2\left(\frac{\kappa^2(2+2i\kappa+\kappa^2)}{32(i+\kappa)(2i+\kappa)}\right)\right.\notag\\
    \left.+(1-u)^3\left(\frac{i\kappa^2(-56+48i\kappa-2\kappa^2+6i\kappa^3+\kappa^4)}{384(i+\kappa)(2i+\kappa)(3i+\kappa)}\right)+\dots\right]\;.
    \label{eq:whole-series-solution-H}
\end{align}
From this approximate solution we can extract a numerical value for $Z$ and its first derivative $Z'$ at any point very close to the horizon. These initial conditions, together with the above equation of motion, define an initial value problem. We can then implement a multiple shooting method that allows us to construct a numerical solution over the whole interval between the horizon and the boundary, $u\in(0,1)$.

A similar analysis can be done near the boundary, where the two series solutions acquire the form
\begin{align}
    \mathcal Z_1(u)&=u^2 F_B(u)\;,\label{eq:ZB1}\\
    \mathcal Z_2(u)&=\mathcal C\log(u)\mathcal Z_1+G_B(u)\;,\label{eq:ZB2}
\end{align}
with $F_B(u)$ and $G_B(u)$ two power series about $u=0$ and $\mathcal C$ a constant. While for arbitrary momentum this constant in general does not vanish, in the particular case of light-like momentum we find $\mathcal C=0$. These functions $F_B(u),\;G_B(u)$, are determined by introducing the expansions~\eqref{eq:ZB1} and~\eqref{eq:ZB2} into the ODE~\eqref{eq:ODE-Z3}. A power series analysis close to the boundary yields
\begin{align}
    \mathcal Z_1&=u^2\left[1+\frac{u^2}{2}-\frac{u^3\kappa^2}{15}+\frac{u^4}{3}-\frac{u^5 19\kappa^2}{210}+\frac{u^6\left(180+\kappa^4\right)}{720}+\dots\right]\;,\label{eq:mathcal-Z1}\\
    \mathcal Z_2&=1-\frac{u^3\kappa^2}{3}-\frac{u^5 4\kappa^2}{15}+\frac{u^6\kappa^4}{72}+\dots\label{eq:mathcal-Z2}
\end{align}
As before, these solutions are used to define an initial value problem and to implement a multiple shooting method to construct a numerical solution over the whole interval between the horizon and the boundary.

Since the ODE~\eqref{eq:ODE-Z3} is second order, the three solutions that we have discussed are not independent. The in-falling solution $Z_1$ (and its canonical momentum) can be expanded in the basis of the two local solutions at boundary $\mathcal Z_1,\mathcal Z_2$, which implies
\begin{subequations}
\begin{align}
    Z_1(u_0)=A\mathcal Z_2(u_0)+B\mathcal Z_1(u_0)\;,\\
    Z'_1(u_0)=A\mathcal Z'_2(u_0)+B\mathcal Z'_1(u_0)\;,
\end{align}
\label{eq:lin-comb-Z}%
\end{subequations}
for any point $u_0$ between the boundary and the horizon. We can therefore use the numerically computed solutions to determine  the connection coefficients $A$ and $B$ for different values of $\kappa$.

\begin{figure}
    \centering
    \includegraphics[width=0.8\linewidth]{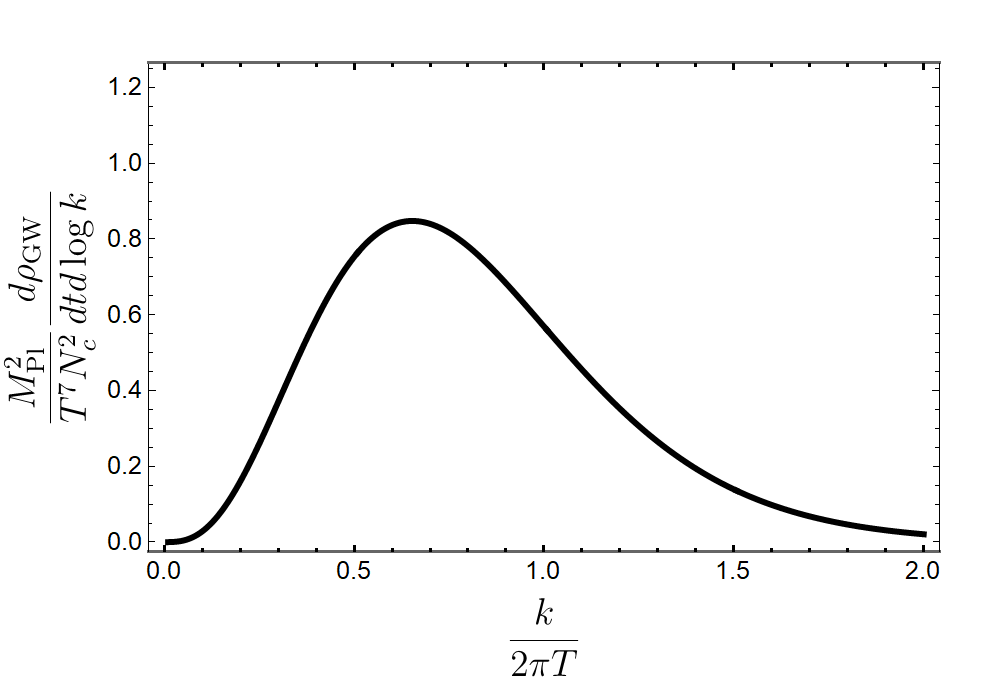}
    \caption{Energy production rate of gravitational waves as a function of dimensionless frequency $\kappa$ in $\mathcal N=4$ SYM at strong $\lambda$.}
    \label{fig:production-rate-figure}
\end{figure}

Once $A$ and $B$ are known, the retarded correlator at light-like momentum condition is given by \cite{kovtun2005quasinormal}
\begin{equation}
    G_3(\kappa)=-\pi^2 N_c^2 T^4\;\frac{B(\kappa)}{A(\kappa)}\;,
    \label{eq:SCSYM-G3}
\end{equation}
with the constant coming from the prefactor $1/16\pi G_5$ of the action~\eqref{eq:SCSYM-action}. However, as discussed in Section~\ref{gauge-string-duality}, in order to implement this result into the computation of the gravitational-wave spectrum, we need to apply the KMS relations and express the retarded propagator in terms of the Wightman function. The relation between both thermal correlators allows us to compute the energy production rate per logarithmic frequency interval by simply introducing Eq.~\eqref{eq:SCSYM-G3} into Eq.~\eqref{eq:log-production-rate-G3}:
\begin{equation}
    \frac{d\rho}{dtd\log k}=\frac{4T^4 k^3}{M_\text{Pl}^2}\;\hat\eta_\text{SCSYM}(k/T)
    \label{eq:final-result-section-4}
\end{equation}
for Planck mass $M_\text{Pl}^2=1/G$, where we have defined a dimensionless function 
\begin{equation}
    \hat\eta_\text{SCSYM}(k/T)=\frac{n_B N_c^2}{2}\Im\frac{B(k/2\pi T)}{A(k/2\pi T)}\;,
    \label{eq:dimensionless-source-term}
\end{equation}
which we shall denote as the source term. The energy spectrum with respect to the dimensionless frequency of the waves is shown in Fig.~\ref{fig:production-rate-figure}. In next sections, we will extend the discussion about this result and its cosmological implications.

\section{Thermal emission rate at weak coupling}\label{comparison-weakly-coupled-source}
So far in this paper, we have derived the thermal spectrum produced by a strongly-coupled thermal source. In this section we will focus on the opposite limit, that is, small interaction between the plasma constituents $\lambda \rightarrow 0$. Even though perturbative analysis has already been made within thermal emission in different gauge theories~\cite{ghiglieri2015gravitational,ghiglieri2020gravitational,ringwald2021gravitational}, if we want to estimate the effect of the coupling on the shape of the GW spectrum it is important to compare the strong and weak coupling computations in the same theory. For this reason, in this section we will determine the thermal production rate of gravitational waves in $\mathcal N=4$ SYM theory at weak 't Hooft coupling $\lambda\rightarrow 0$ (WCSYM), while keeping the large $N_c\rightarrow \infty$ approximation.

For our perturbative analysis, we will take advantage of the generalization of the perturbative calculations of~\cite{ghiglieri2015gravitational,ghiglieri2020gravitational} to a theory with arbitrary gauge group, Weyl fermion and scalar content performed in~\cite{ringwald2021gravitational}. As it is well known, computing thermal correlators in perturbation theory for all possible values of the external momentum is a complicated task that requires different approximations for different momentum regions. Here, as in~\cite{ghiglieri2015gravitational,ghiglieri2020gravitational,ringwald2021gravitational}, we will only focus on the region where the momentum of the correlator is much larger than the Debye thermal mass, $k\gtrsim m(T)$, since it is in this region where the perturbative thermal emission rate is maximal. In this momentum range, the complete leading-order result of~\cite{ghiglieri2020gravitational} for the spin-2 projection of the energy-momentum tensor\footnote{The analysis in~\cite{ghiglieri2020gravitational} did not include the so-called ``improvement terms" of the energy-momentum tensor~\cite{Callan:1970ze}, which arise as a result of the coupling of scalar fields to the Ricci tensor. Those terms have an important effect on the radiation pattern by an external source~\cite{Fiol:2019woe}. However, they are either longitudinal or vanish for light-like momenta, and therefore do not contribute to $T^\text{TT}_{ij}$.} 
two point function appearing in Eq.~\eqref{eq:production-rate} was extended in~\cite{ringwald2021gravitational} for a general $SU(N_c)$ gauge theory\footnote{The analysis in~\cite{ringwald2021gravitational} allowed arbitrary gauge groups. We restrict their analysis to $SU(N_c)$ theories to lighten the discussion.}
to give
\begin{align}
    \hat\eta\left(\frac{k}{T}\equiv\hat k\right)\simeq&\;\hat\eta_\text{HTL}(\hat k)\notag\\
    &+g^2(N_c^2-1)\left(\frac{1}{2}T_\text{Ad}\eta_{gg}(\hat k)+\sum_i T_i\eta_{sg}(\hat k)+\frac{1}{2}\sum_\alpha T_\alpha\eta_{fg}(\hat k)\right)\notag\\
    &+\frac{1}{4}\sum_{i\alpha\beta}|y_{\alpha\beta}^{iabc}|^2\eta_{sf}(\hat k)\;,
    \label{eq:dimensionless-function-eta}
\end{align}
where  $T_\text{Ad}=N_c$, $T_i$ and  $T_\alpha$ are the Dynkin indices of the Adjoint representation of the gauge group, and of the individual gauge  representations of the scalar and fermion fields, respectively. The four two-loop functions $\eta_{gg}(\hat k)$, $\eta_{sg}(\hat k)$, $\eta_{fg}(\hat k)$, $\eta_{sf}(\hat k)$,  involve diagrams with only gauge fields, scalar and gauge fields, fermions and gauge fields, and scalars and fermions, respectively (their full description and mathematical definition can be found in~\cite{ghiglieri2020gravitational}). In addition, the ``hard thermal loop'' (HTL) resummation function $\hat\eta_\text{HTL}$ is given by
\begin{equation}
    \hat\eta_\text{HTL}(\hat k)=\frac{\hat k}{16\pi(e^{\hat k}-1)}(N_c^2-1)\hat m^2\log\left(1+4\frac{\hat k^2}{\hat m^2}\right)\;,
\end{equation}
which depends on the Debye screening length of the plasma, given at leading order by
\begin{equation}
    m^2(T)=g^2 T^2\left(\frac{1}{3}T_\text{Ad}+\frac{1}{6}\sum_{i}T_{i}+\frac{1}{6}\sum_{\alpha}T_{\alpha}\right)\equiv\hat m^2 T^2\;.
    \label{eq:mDdef}
\end{equation}
Finally, $\hat\eta$ depends on the Yukawa couplings among fermions and scalars, which in~\cite{ringwald2021gravitational} are parametrized as 
\begin{equation}
    \mathcal L_\text{Yukawa}= -  \sum_{i, \alpha, \beta}  y_{\alpha\beta}^{i} \phi_i \psi_\alpha \psi_\beta + \rm{h. c.}\,,
    \label{eq:LYukawagen}
\end{equation}
where $\phi$ and $\psi$ are the scalar and Weyl fermion fields of the theory. In this contribution of the Lagrangian, all internal indices are traced over.

To obtain the thermal emission rate in $\mathcal N=4$ SYM, we particularize those general results to this precise case. This theory contains four Weyl fermions $\psi_\alpha,\;\alpha=1,\dots,4$, and six real scalar fields $\Phi_i,\;i=1,\dots,6$. Both these two types of matter fields are in the adjoint representation of the gauge group, and therefore their Dynkin indices are $T_i=T_\alpha=T_\text{Ad}=N_c$. Introducing these properties of the matter fields into the general expression for the Debye mass~\eqref{eq:mDdef}, we obtain the leading-order expression $\hat m^2_\text{WCSYM}= 2\lambda$, with $\lambda$ the 't Hooft coupling. This knowledge of the matter content is sufficient to determine all contributions to the production rate in Eq.~\eqref{eq:dimensionless-function-eta} but those proportional to the Yukawa couplings. The $\N=4$ SYM  Lagrangian does not possess a term of the form~\eqref{eq:LYukawagen}; however, it does possess interacting terms involving two fermions and one scalar field. To write this term explicitly, we use the representation of the  $\N=4$ Lagrangian used in~\cite{Yamada:2006rx,du2020two}, where the scalar fields are expressed as a multiplet
\begin{equation}
    \Phi\equiv(X_1,Y_1,X_2,Y_2,X_3,Y_3)\;,
\end{equation}
where $X_p$ and $Y_q$, $p,q=1,2,3$, denote scalar and pseudoscalar fields. The interaction Lagrangian of the scalar with two fermions is then 
\begin{equation}
    \mathcal L_{\Phi \psi \psi}=\Tr{-ig\bar\Psi_\alpha\comm{A_{\alpha\beta}^p X_p+iB_{\alpha\beta}^q \gamma_5 Y_q}{\Psi_\beta}}\;,
    \label{eq:LPhipsipsi}
\end{equation}
where  we have introduced a Majorana fermion $\bar \Psi_\alpha\equiv \left(\psi_\alpha, \bar \psi_\alpha\right)$ and $A^p$ and $B^q$ are $4\times 4$ matrices that satisfy $A_{\alpha\gamma}^p A_{\gamma\beta}^p=B_{\alpha\gamma}^p B_{\gamma\beta}^p=-3\delta_{\alpha\beta}$. The explicit form of these matrices is given in terms of the $2\times 2$ Pauli matrices as
\begin{align}
    A^1&=\begin{pmatrix}0&\sigma_1\\-\sigma_1&0\end{pmatrix}\;,\quad
    A^2=\begin{pmatrix}0&-\sigma_3\\\sigma_3&0\end{pmatrix}\;,\quad
    A^3=\begin{pmatrix}i\sigma_2&0\\0&i\sigma_2\end{pmatrix}\;,\notag\\
    B^1&=\begin{pmatrix}0&i\sigma_2\\i\sigma_2&0\end{pmatrix}\;,\quad
    B^2=\begin{pmatrix}0&\sigma_0\\-\sigma_0&0\end{pmatrix}\;,\quad
    B^3=\begin{pmatrix}-i\sigma_2&0\\0&i\sigma_2\end{pmatrix}\;.
\end{align}
The main difference between the interaction Lagrangian~\eqref{eq:LPhipsipsi} and the Standard-Model-like Yukawa Lagrangian~\eqref{eq:LYukawagen} lies in the gauge-index structure of the interaction. However, for leading-order computation, the gauge indices in Eq.~\eqref{eq:LPhipsipsi} may be regarded as internal indices, since at this order the fields interacting via this vertex do not suffer color exchanges. Expanding the matter fields in the basis of color generators as $\Phi_i=\Phi_i^a t^a$ and $\psi_\alpha=\psi_\alpha^a t^a$, with $t^a$ the $SU(N_c)$ generators in the fundamental representation,\footnote{The normalization of the generators is fixed by the condition $\rm{Tr}\left(t^a t^b\right)=\frac{1}{2} \delta^{a b}$.}
we can identify
\begin{equation}
    \mathcal L_{\Phi \psi \psi}\equiv y_{\alpha\beta}^{iabc}\phi_i^b\psi_\alpha^a\psi_\beta^c\;,
 \label{eq:ydef}
\end{equation}
where the couplings $y_{\alpha\beta}^{iabc}$ are combinations of the $A$ and $B$ matrices and the structure constants of $SU(N_c)$. Using these relations, we find 
\begin{align}
\sum_{i\alpha\beta}|y_{\alpha\beta}^{iabc}|^2&=\frac{g^2}{4} f^{abc} f^{abc}  \left( A^p_{\alpha \beta} A^{*p}_{\alpha \beta} +B^p_{\alpha \beta} B^{*p}_{\alpha \beta}\right)\notag\\
&=6g^2N_c(N_c^2-1)\;.
\end{align}
We have already identified the values of all parameters appearing in the $\mathcal N=4$ SYM source term~\eqref{eq:dimensionless-function-eta}. Taking the large-$N_c$ limit to facilitate the comparison with the strong-coupling calculation, the final leading-order expression for this conformal field theory is 
\begin{align}
    \hat\eta_\text{WCSYM}(\lambda,\hat k)&\simeq\lambda N_c^2\left[\frac{\hat k}{8\pi(e^{\hat k}-1)}\log\left(1+2\frac{\hat k^2}{\lambda}\right)+\frac{1}{2}\eta_{gg}(\hat k)+6\eta_{sg}(\hat k)+2\eta_{fg}(\hat k)+\frac{3}{2}\eta_{sf}(\hat k)\right]
\label{eq:etaweak}
\end{align}
Note that the $\lambda$-dependence of this expression is non-trivial. The two-loop functions $\eta_{gg}$, $\eta_{sg}$, $\eta_{fg}$ and $\eta_{sf}$ depend solely on the dimensionless ratio $\hat k=k/T$ and their contribution to the correlator is proportional to $\lambda$, as expected from a leading-order calculation. However, this is not the case for the $\hat\eta_\text{HTL}$ function. The reason for this non-trivial dependence on the coupling is that this contribution resums a set of higher-order diagrams that become large for momenta $\hat k\sim\hat m$, via the HTL approximation. Note also that an important assumption of the HTL framework is that the temperature and the Debye mass scales are well separated, i.e., $m\ll T$. We therefore expect this result to hold for the $\lambda \ll 1/2$ regime in $\mathcal N=4$ SYM theory.

\begin{figure}
    \centering
    \includegraphics[width=0.8\linewidth]{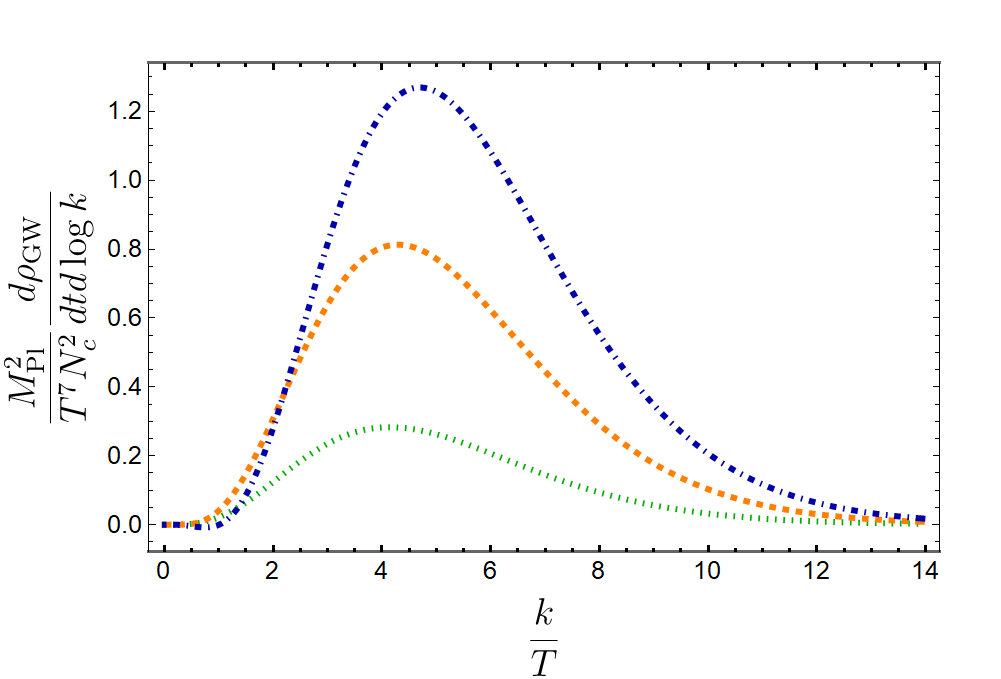}
    \caption{Comparison between the energy density of gravitational waves produced by quasi-particle excitations in a weakly-coupled thermal plasma for different values of the coupling constant $\lambda$. The lines are colored for the values of the wave number in which the computation can be trusted ($\hat k>2\lambda$). The spectra correspond to $\lambda=0.1$ (dotted green), $\lambda=0.5$ (dashed orange) and $\lambda=1.8$ (dot-dashed blue). The peaks of the spectra can be seen to occur around $k\sim T$ for different values of $\lambda$.}
    \label{fig:spectrum-comparison-gray}
\end{figure}

Now that we have introduced all the features and quantities characterizing our theory, we can derive the spectrum of stochastic thermal waves emitted within weakly-coupled $\mathcal N=4$ SYM. Analogously to Eq.~\eqref{eq:final-result-section-4}, the energy density carried by gravitational waves per unit time and unit logarithmic wave number interval is given by
\begin{equation}
    \frac{d\rho}{dtd\log k}=\frac{4T^4 k^3}{M_\text{Pl}^2}\;\hat\eta_\text{WCSYM}(\lambda,\hat k)\;.
        \label{eq:drhopert}
\end{equation}
As we can see, the source term and, consequently, the shape of the spectrum emitted by a weakly-coupled thermal plasma in SYM depend on the coupling constant $\lambda$ of the theory. In order to visualize this coupling-dependence of the emission rate, the full spectra of thermally produced gravitational waves in the perturbative regime is shown in Fig.~\ref{fig:spectrum-comparison-gray} for convenient values of $\lambda$: $\lambda=0.1$ (dotted green), $\lambda=0.5$ (dashed orange) and $\lambda=1.8$ (dot-dashed blue). Note that only the first one of these values satisfies $\hat m<1$ and may therefore lie within the regime of validity of this calculation. Based on these criteria, the second value $\lambda=0.5$ marks in fact the limit beyond which this analysis is no longer valid. Nevertheless, it will also be useful to extrapolate the perturbative calculation to intermediate values of $\lambda$ in order to compare its behaviour with the strong coupling computation.

\begin{figure}
    \centering
    \begin{subfigure}[b]{0.49\textwidth}
        \centering
        \includegraphics[width=\linewidth]{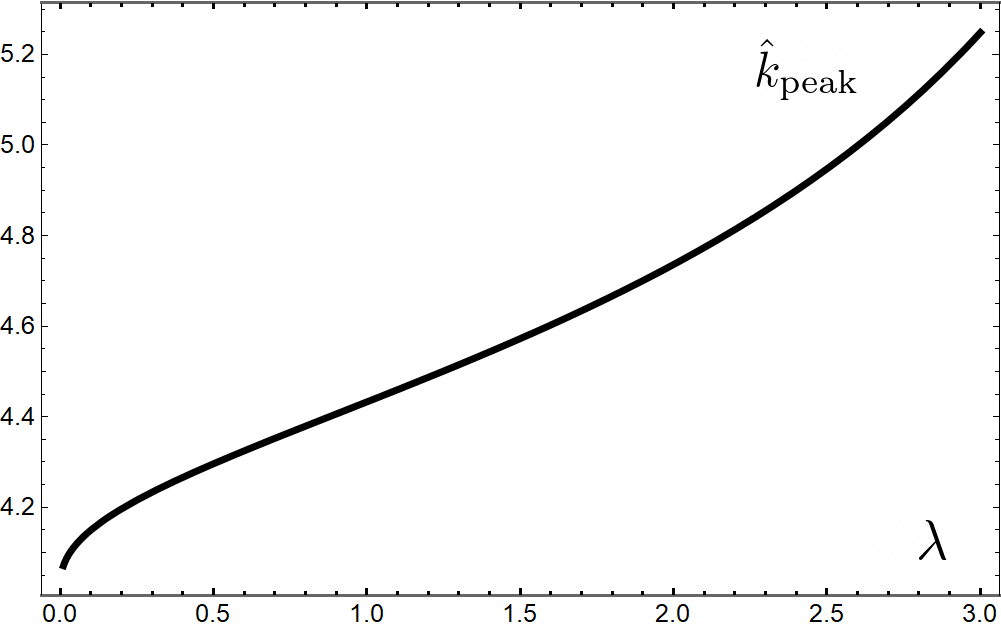}
    \end{subfigure}
    \begin{subfigure}[b]{0.49\textwidth}
        \centering
        \includegraphics[width=\linewidth]{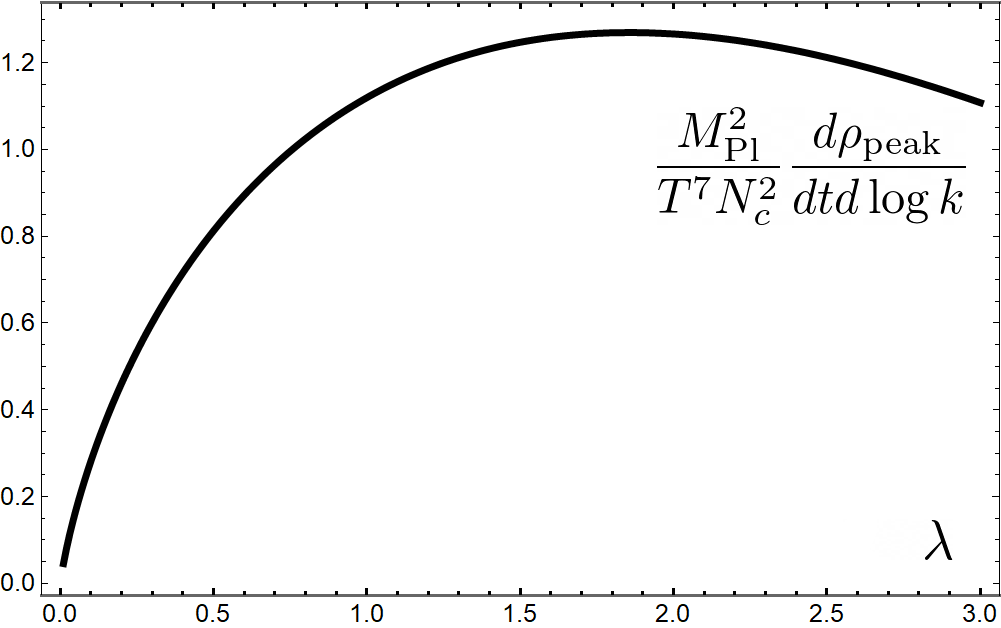}
    \end{subfigure}
    \caption{Left: position of the peak wave number with respect to the coupling constant $\lambda$; the perturbative analysis predicts that an increase in the interaction strength of the thermal plasma shifts the peak energy towards higher frequencies. Right: peak energy density with respect to the coupling constant $\lambda$; contrary to $\hat k_\text{peak}$, the peak energy does not increase monotonically with $\lambda$, but rather increases until reaching a maximum value at $\lambda_\text{max}\sim 1.8$, then decreases.}
    \label{fig:peak-energy-density}
\end{figure}

To better understand the main features of the perturbative calculation, we study the $\lambda$-dependence of two key characteristics of the spectrum: the value of the spectrum at the extrema (peak energy density) and the location of the latter. These two quantities are shown in Fig.~\ref{fig:peak-energy-density}. As we can see in the left panel, the position of the peak increases monotonically with the coupling. On the contrary, the behaviour of the maximum of the spectrum is not monotonic: at small coupling, where the perturbative computation is valid, the energy radiated in gravitational waves increases as the coupling grows. This behaviour is expected, since as the coupling grows the plasma constituents suffer more violent collisions, leading to an increase in the emitted radiation. However, when we extrapolate the perturbative calculation to $\lambda \sim 1$, the maximum of the spectrum starts to decrease, as shown in the right panel of this figure. This maximum in fact occurs at $\lambda\simeq 1.8$; the leading-order perturbative spectrum extrapolated to this value is shown in the dot-dashed blue curve of Fig.~\ref{fig:spectrum-comparison-gray}. We will come back to this non-monotonic behaviour when comparing this calculation with the strong-coupling analysis of Section~\ref{TERSC}.

\section{Cosmological evolution}\label{SYM-in-cosmology}

The analysis we have performed so far within the framework of $\mathcal N=4$ SYM theory has been restricted to the emission from a static thermal source. However, in any conceivable situation in which the thermally-emitted gravitational waves might be detected, matter does not remain static but expands with the expansion of the Universe. Certainly, no form of matter in the Universe is governed by $\N=4$ SYM. Nevertheless, if we want to use the $\N=4$ calculations to build intuition about the coupling-dependence of the emission rate, we must include the cosmological history in our analysis, since this process strongly modifies the magnitude and position of the gravitational-wave signal.

For this study, we will assume that over a certain time in the early, very hot history of the Universe, matter dynamics could be understood in terms of $\N=4$ SYM interactions. We will also assume this period to extend from a very early time $t_0$ when the temperature of the Universe was $T_{\rm max}$ to a later time $t_{\rm end}$ with much lower temperature $T_{\rm end}\ll T_{\rm max}$. Later than that, matter dynamics stops being close to $\N=4$ SYM. Even though matter may still be interacting below $T_{\rm end}$, we will assume the output of gravitational waves to be negligible for times larger than $t_{\rm end}$. This assumption is justified since the spectrum of gravitational waves scales with the temperature of the plasma~\cite{ringwald2021gravitational}. Therefore, the energy emitted by thermalized matter is dominated by the range $T_{\rm end}< T< T_{\rm max}$. In this region of temperatures, the expansion of the Universe is also taken to be dominated by the energy density of the thermalized matter $e(T)$, that is,
\begin{equation}
    H=\frac{\dot a}{a}=\sqrt{\frac{8\pi e}{3}}\frac{1}{M_\text{Pl}}\;. 
    \label{eq:hubble-scale-factor}
\end{equation}
This expansion rate implies that the temperature of the system decreases with time. Since the theory we are considering is conformal, the ratio $T(t)/a(t)$ remains constant, leading to the following relation between time and temperature:
\begin{equation}
\frac{1}{T}\frac{d T}{ dt} = - \sqrt{\frac{8\pi e}{3}}\frac{1}{M_\text{Pl}}\;.
\label{eq:dTdtgen}
\end{equation}
This expression may be further simplified by exploiting the conformal invariance. Following the notation in Ref.~\cite{ringwald2021gravitational}, we define the effective number of degrees of freedom of the theory as 
\begin{equation}
    \quad g_{*s}(T)=\frac{s(T)}{\frac{2\pi^2}{45}T^3}\;,
    \label{eq:e-s-c-densities}
\end{equation}
where $s$ accounts for the entropy density. In general, $g_{*s}(T)$ is a temperature-dependent function. However, for the particular case of a conformal theory, it becomes a coupling-dependent constant. Conformal invariance also implies that the energy density is given by $e= 3 g_{*s} T^4/4$. Therefore, in a conformal theory, the relation~\eqref{eq:dTdtgen} becomes
\begin{equation}
    \dv{T}{t}=-\frac{\pi}{\sqrt{90}}\;g_{*s}^{1/2}\;\frac{T^3}{M_\text{Pl}}\;.
    \label{eq:dTdtconf}
\end{equation}
For $\N=4$ SYM in the large-$N_c$ limit, the effective number of degrees of freedom in the $\lambda\rightarrow 0 $ limit may be obtained directly from the matter content of the theory and is given by\footnote{Two-loop HTL-resummed corrections to this quantity can be found in Ref.~\cite{du2020two}. 
} $g_{*s,\text{WC}}=15N_c^2$. Meanwhile, for the $\lambda \rightarrow \infty$ limit, the holographic computation of the entropy~\cite{Gubser:1996de} yields $g_{*s,\text{SC}}=\frac{45}{4}N_c^2$.

The cosmological expansion characterized by $H$ induces a dilution of the energy density carried by gravitational waves. As long as $H\ll T$, we can identify the microscopic calculation at fixed temperature as a rate of energy injection per unit volume to this component of the energy density. As the Universe cools down, this rate decreases as time progresses. To simplify the discussion, in this section we will denote by $R$ the energy spectrum of gravitational waves per unit three momentum which, as we know, is a function of the coupling constant in $\mathcal N=4$ SYM. Compactifying Eqs.~\eqref{eq:final-result-section-4} and~\eqref{eq:drhopert} into one single expression:
 \begin{equation}
    R(T,\lambda,\hat k)\equiv \frac{4T^4}{M_\text{Pl}^2}\;\hat\eta(\hat k)= \left\lbrace\begin{array}{l}\frac{4T^4}{M_\text{Pl}^2}\;\hat\eta_\text{WCSYM}(\lambda,\hat k)\;,\phantom{abcd}2\lambda\lesssim 1\;,\\ \frac{4T^4}{M_\text{Pl}^2}\;\hat\eta_\text{SCSYM}\left(\frac{\hat k}{2\pi}\right)\;,\phantom{abcd}\lambda\rightarrow\infty\;.\end{array}\right.
    \label{eq:R-WC-SC}
\end{equation}
The energy injected by this rate adds to the total energy density of gravitational waves, which, taking into account the expansion rate, satisfies the differential equation~\cite{ghiglieri2015gravitational}
\begin{equation}
    \left(\partial_t+4H\right)\rho_\text{GW}(t)=\int_\mathbf{k}\frac{d^3\mathbf k}{(2\pi)^3}\;R(T,\lambda,\hat k)\;.
    \label{eq:evolution-equation}
\end{equation}
This differential equation may be easily integrated from $t_0$ to $t_\text{end}$. This way, we can express the energy density of gravitational waves at $t_{\rm end}$ as 
\begin{align}
    \rho_\text{GW}(t_\text{end}) &= s^{4/3}(t_\text{end}) \int_{t_\text{0}}^{t_\text{end}}d t\int_\mathbf{k}\frac{d^3\mathbf k}{(2\pi)^3}\frac{R(T,\lambda,\hat k)}{s^{4/3}(t)}\notag\\
    &=\frac{\sqrt{90}\;M_\text{Pl}}{\pi}  \frac{T_\text{end}}{g_{*s}^{1/2}}\int_{T_\text{end}}^{T_\text{max}}\frac{d T}{T^4}\int_\mathbf{k}\frac{d^3\mathbf k}{(2\pi)^3}\;R(T,\lambda,\hat k)\;,
\end{align}
where in the second line we have used the relation~\eqref{eq:dTdtconf} and the definition of the effective number of degrees of freedom~\eqref{eq:e-s-c-densities}.

In addition to diluting the energy density, the expansion of the Universe also changes the frequency of the emitted waves. Therefore, to determine the spectrum at a fixed value of  $k_\text{end}$, the wave momentum at $t_\text{end}$, and since  momenta redshift as $k(t)=k_\text{end}\;a(t_\text{end})/a(t)$, we must evaluate the emission rate at different values as time changes. However, in a conformal theory, the ratio $k/T$ remains constant during the expansion. Taking this into account, the GW energy density per logarithmic wave number interval at $t_\text{end}$ turns into
\begin{align}
    \Omega_\text{end}(k_\text{end})\equiv&\;\frac{1}{e(T_\text{end})}\frac{d\rho_\text{GW}}{d\log k_\text{end}}(T_\text{end},k_\text{end})\notag\\
    =&\;\frac{720\sqrt{10}\;M_\text{Pl}}{2\pi^2(2\pi)^3}\frac{1}{g^{3/2}_{*s}}\frac{k_\text{end}^3}{T_\text{end}^3}
    \int_{T_\text{end}}^{T_\text{max}}\frac{d T}{T^4}\;R\left(T,\lambda,\frac{k_\text{end}}{T_\text{end}}\right)\;,\label{eq:Omega-ewco}
\end{align}
where we have introduced the identity~\eqref{eq:e-s-c-densities}. Using Eq.~\eqref{eq:R-WC-SC}, we observe that the integrand in this expression does not depend on $T$, and the integral can be easily performed. Assuming $T_\text{max} \gg T_\text{end}$ we neglect the lower limit of integration and express
\begin{equation}
    \Omega_\text{end}(k_\text{end})=\frac{1440\sqrt{10}\;}{\pi^2(2\pi)^3 }\frac{1}{g^{3/2}_{*s}}\frac{k_\text{end}^3}{T_\text{end}^3} \frac{T_\text{max}}{M_\text{Pl}}\;\hat\eta \left(\frac{k_\text{end}}{T_\text{end}}\right)\;.
\end{equation}

Although we assume that the production of gravitational waves below $T_\text{end}$ is negligible, the expansion of the Universe continues redshifting and diluting the spectrum. The effective number of degrees of freedom also changes after $t_\text{end}$, since at the end of the evolution the Universe contains the current matter content. We will assume that these changes are fast and are followed by long periods in which the equation of state is  approximately conformal. With this picture in mind, we can approximate the dilution of the spectrum from $t_\text{end}$ up to the final time $t_\text{fin}$ by $\rho(\text{fin})\simeq\rho_\text{end}\;a^4(t_\text{end})/a^4(\text{fin})$, which implies 
\begin{align}
    \Omega(k_\text{end})\equiv&\;\frac{1}{e_\text{cr}}\dv{\rho_\text{GW}(\text{fin})}{\log k_\text{end}}\notag\simeq\frac{1}{e_\text{cr}}\left(\frac{s(\text{fin})}{s(T_\text{end})}\right)^{\frac{4}{3}}e(T_\text{end})\;\Omega_\text{end}(k_\text{end})\notag\\
    =&\;\frac{720\sqrt{10}}{\pi^2(2\pi)^3}\;\Omega_\gamma\;\frac{g_{*s}^{4/3}(\text{fin})}{g^{11/6}_{*s}}\frac{k_\text{end}^3}{T_\text{end}^3}  \frac{T_\text{max} } {M_\text{Pl}}\;\hat \eta \left(\frac{k_\text{end}}{T_\text{end}}\right)\;,
\end{align}
where $g_{*s}(\text{fin})=3.931\pm 0.004$~\cite{Saikawa:2018rcs} is the number of entropy degrees of freedom after neutrino decoupling and
\begin{equation}
    \Omega_\gamma\equiv\frac{e_\gamma(\text{fin})}{e_\text{cr}(\text{fin})}=\frac{2\pi^2}{30}\frac{T_0^4}{e_\text{cr}}=2.4728(21)\times 10^{-5}/h^2
\end{equation}
is the present energy density of the CMB photons, with temperature $T_0=2.72548(57)$ K.

Finally, we need to express the spectrum in terms of the present-day frequency. Assuming once again close-to-conformal expansion, the current-day frequency of the waves is related to $k_\text{end}$ as
\begin{equation}
    f=\frac{1}{2\pi}\frac{a(t_\text{end})}{a(\text{fin})}k_\text{end}\;,
\end{equation}
leading to a suitable expression
\begin{equation}
    \frac{k_\text{end}}{T_\text{end}}=2\pi\left(\frac{g_{*s}}{g_{*s}(\text{fin})}\right)^{\frac{1}{3}}\frac{f}{T_0}\;.
\end{equation}
Using these relations, the present-day spectrum is given by 
\begin{align}
    &\Omega(f)\simeq\frac{1440\sqrt{10}}{2\pi^2}\;\Omega_\gamma\;\frac{g_{*s}^{1/3}(\text{fin})}{g_{*s}^{5/6}}\frac{f^3}{T_0^3}\frac{T_\text{max}}{M_\text{Pl}}\;\hat\eta\left(2\pi\left[\frac{g_{*s}}{g_{*s}(\text{fin})}\right]^{\frac{1}{3}}\frac{f}{T_0}\right)\;.\label{eq:current-spectra}
\end{align}
Substituting either the strong coupling~\eqref{eq:dimensionless-source-term} or the weak coupling source term~\eqref{eq:etaweak} into this expression, we can draw a numerical estimate of the gravitational waves energy production rate in the context of an expanding cosmological model for the different values of the coupling constant. Note that both in the weak and strong coupling calculations the number of effective degrees of freedom is proportional to $N_c^2$. The magnitude and the position of the peak energy density are thus expected to depend on the number of colours of the theory, which we do not specify. This sets an additional degree of freedom in the shape of the spectrum. Finally, one should also notice that $g_{*s}$ at strong and weak coupling differ by a $3/4$ factor, which shifts the spectrum to somewhat different frequencies as a consequence of the expansion.

\begin{figure}
    \centering
    \begin{subfigure}[b]{0.8\textwidth}
        \centering
        \includegraphics[width=\linewidth]{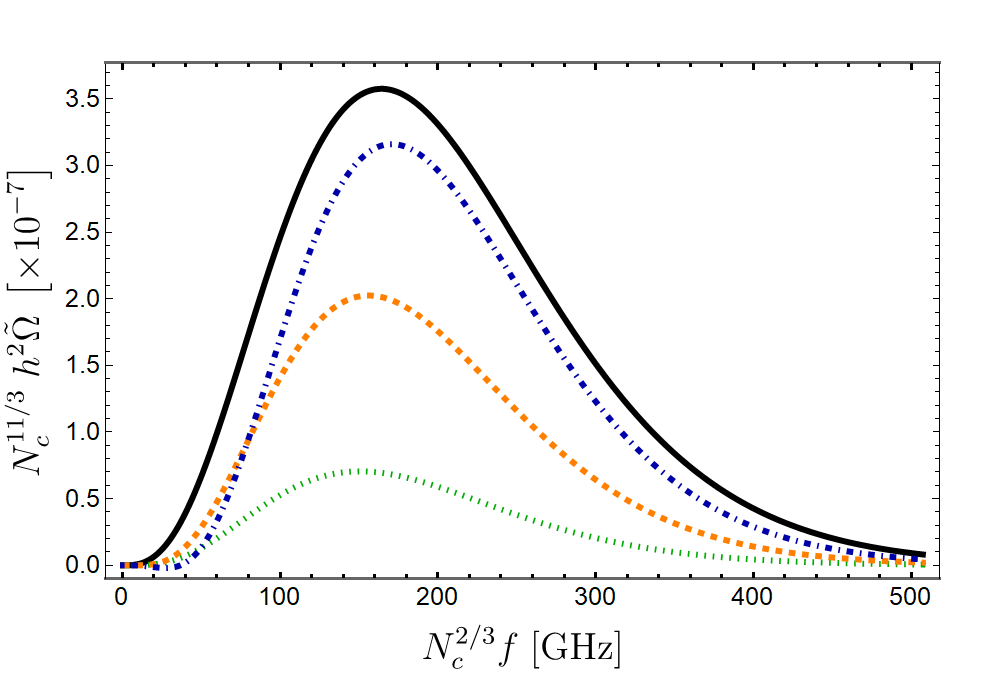}
    \end{subfigure}
    \begin{subfigure}[b]{0.8\textwidth}
        \centering
        \includegraphics[width=\linewidth]{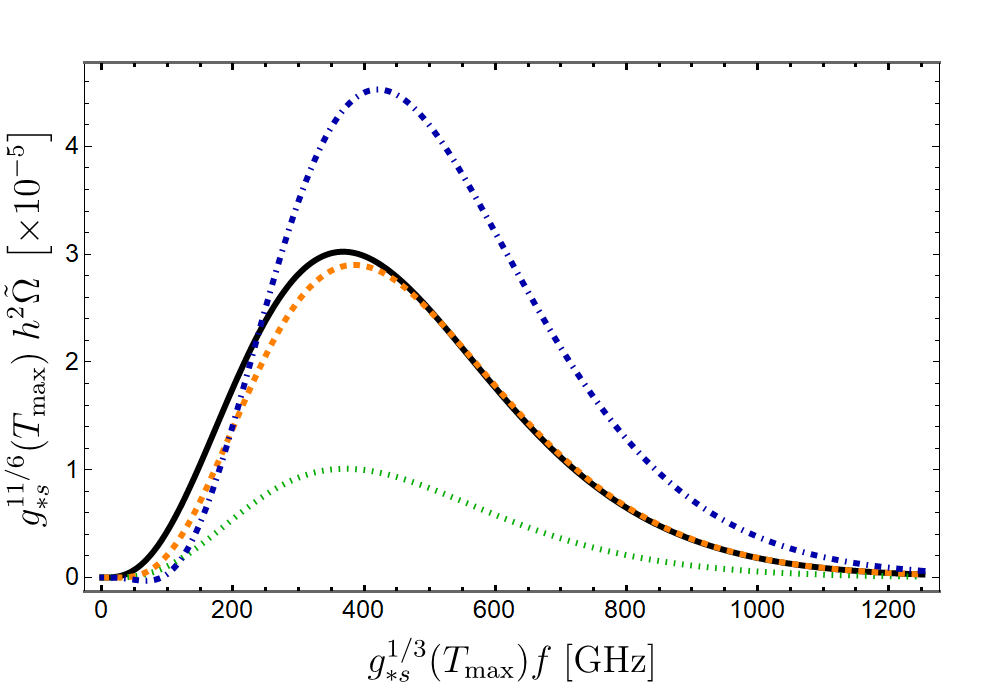}
    \end{subfigure}
    \caption{Gravitational waves energy density as a function of the current frequency $f$ after convolution with the expansion of the Universe, with $\tilde\Omega\equiv\frac{M_\text{Pl}}{T_\text{max}}\frac{\Omega}{N_c^2}$. The spectra correspond to the (holographic) strong coupling calculation (solid black), $\lambda=0.1$ (dotted green), $\lambda=0.5$ (dashed orange) and $\lambda=1.8$ (dot-dashed blue). The peak of the spectra occur around $N_c^{2/3}f\sim 150$ GHz for different values of $\lambda$.}
    \label{fig:spectrum-comparison}
\end{figure}

The result of these two extreme calculations are shown in the top panel of  Fig.~\ref{fig:spectrum-comparison}. In this plot, the strong coupling result corresponds to the solid black line while the weak coupling calculations for $\lambda=0.1,\, 0.5,\, 1.8$ correspond to the dotted green, dashed orange and dot-dashed blue, respectively. Note that, for all the values of the coupling, the strong coupling result has a larger magnitude in the spectrum of gravitational waves. Since, as we have already seen, $\lambda=1.8$ corresponds to the maximum value of the peak energy density we can obtain from the extrapolation of the perturbative, leading-order result, we can ensure that the strong coupling calculation produces more gravitational waves than the extrapolated perturbative calculation for any value of the coupling. Note, however, that the main features of the strong coupling spectrum are well-captured by the maximum perturbative result at $\lambda=1.8$.

This qualitative agreement between the strong coupling and $\lambda=1.8$ computations is, in fact, the result of two different contributions. On the one hand, the thermal source $\hat\eta$ at the peak position is larger for the perturbative result than for the strong coupling calculation, as it may be inferred from inspection of Fig.~\ref{fig:production-rate-figure} and Fig.~\ref{fig:spectrum-comparison-gray}. On the other hand, the effective number of degrees of freedom is reduced at strong coupling. This enhances the strong coupling spectrum since, at equal temperature, the smaller number of degrees of freedom slows the expansion and leads to a smaller dilution of the signal.

To isolate the effect of the microscopic emission rate, in the lower panel of Fig.~\ref{fig:spectrum-comparison} we have scaled-out the number of degrees of freedom from both the frequency and the magnitude of the spectrum. By doing this scaling, the final spectra become a redshift of the thermal spectra of Fig.~\ref{fig:production-rate-figure} and Fig.~\ref{fig:spectrum-comparison-gray} together with a reduction of the amplitude common to all calculations. After this scaling, the extrapolated $\lambda=1.8$ result is no longer a good approximation to the strong coupling result. However, after the degrees of freedom are factored out, we find a surprising semi-quantitative agreement between the strong coupling result and the perturbative extrapolation to $\lambda=0.5$. Note that this value marks a limit of applicability of the perturbative analysis, since beyond this value the Debye mass is not separated from the $T$ scale. This means that $\lambda=0.5$  may be used as a good predictor of the strong coupling limit, at least in a static thermal system.

\section{Discussion}\label{disc}
The spectrum of thermally-emitted gravitational waves is an interesting observable which could potentially provide us with valuable information of the very hot stages of the Universe, which are hardly accessible to other probes. Nevertheless, this is also a very challenging observable, since the region of frequencies where the spectrum is maximal, in the vicinity of 100 GHz, is not within our current detection capabilities. Nevertheless, the interest in high-frequency gravitational waves is pushing the community to devise new detection technologies. Therefore, it is worth analyzing this mechanism of GW production in depth.

In this paper we have used $\N=4$ SYM as a tool with which to understand this process. Certainly, this theory is very different from the Standard Model and modelling the matter present in the universe with this theory is in no possible sense an approximation. However, this theory allows us to determine the gravitational-wave spectrum in very different dynamical regimes. In particular, and thanks to the gauge/gravity duality, we can access the plasma dynamics at strong coupling, when the plasma cannot be considered as a collection of interacting quasi-particles, in contrast to perturbation theory. This fact implies that calculations based on $\N=4$ may be used not only to get an intuition on how the emission rate depends on a particular parameter of the theory, but also whether this observable is sensitive to the presence or absence of quasi-particles in the plasma. There is little doubt that unbroken phase of the Standard Model is dominated by quasi-particles. However, thermal dynamics at very large scales or even in the dark matter sector, if it behaves thermally at some early stage of the evolution of the Universe, could have a strongly-coupled phase with no quasi-particles. The strategy to use the gauge/gravity duality to describe the dynamics at strong coupling has been very successful in understanding the properties of the QCD plasma close to the deconfinement transition (see~\cite{casalderrey2014gauge} for an extended review and~\cite{Caron-Huot:2006pee} for an analysis of thermal photon and dilepton emission, which are weakly-coupled probes of the QCD plasma, similar to the one described in our work). For other applications of the gauge/gravity duality to understand different processes of gravitational-wave emissions see~\cite{Bea:2021zol,Bea:2021zsu,Ares:2020lbt,Ares:2021ntv,Ares:2021nap,Bigazzi:2020avc,Bigazzi:2021ucw}.

In this work, we have focused on the spectrum of gravitational waves produced by a thermal system at momenta in the vicinity of the maximum of the emission rate. At very small momenta, the emission rate is solely controlled by the shear viscosity of the system~\cite{ghiglieri2015gravitational}. Since viscosity is dominated by the least interacting degrees of freedom~\cite{Arnold:2000dr}, we have not explored this region with our strong coupling computation. In addition, the rate of gravitational-wave emission is extremely small for those small frequencies, much smaller than other known sources of stochastic gravitational waves~\cite{ghiglieri2015gravitational,ghiglieri2020gravitational,ringwald2021gravitational}. On the contrary, the maximum of the spectrum is more sensitive to interacting degrees of freedom. As our calculation shows, the qualitative features of the thermally-produced gravitational-wave spectrum are very similar for the two extreme microscopic pictures of the plasma constituents described by the $\lambda\rightarrow 0$ and $\lambda \rightarrow\infty$ limits of $\N=4$ SYM. Nevertheless, there are quantitative differences between those two extremes.

For very weakly-coupled plasmas, the emission rate of gravitational waves is suppressed by the smallness of the coupling, since this radiation cannot be produced in the absence of matter interaction. Naively, the perturbative spectrum is proportional to $\lambda$, but the complicated infrared dynamics of gauge theory plasmas alters this behaviour as a consequence of the emergence of the Debye mass scale. As the coupling increases, so does the emission rate. However, it is hard to assess whether this growth should be monotonous with the coupling constant. Nevertheless, the strong coupling computation provides a natural benchmark to estimate the maximum strength of the emission rate, at least for $\N=4$ SYM. In this paper, we have compared this limit with extrapolation of the leading-order perturbative calculation to intermediate values of the coupling. Even though next-to-leading order correction should be significant for the large values of the coupling we have considered, these results provide us with a second reference to assess the spectrum.

Remarkably, the extrapolation of the perturbative spectrum to intermediate coupling is not monotonous but rather possesses a maximum at $\lambda\simeq 1.8$. Because of this, leading-order perturbative computations in this region show very little sensitivity to the particular value of the coupling. At fixed temperature, this maximum of the perturbative emission rate leads to larger output of gravitational waves than the strong coupling calculation at comparable momenta. However, if the fixed temperature computations are embedded into a cosmological expansion model, these two calculations yield a comparable spectrum of gravitational waves at present time. Nevertheless, this observation depends significantly on the change of the effective number of degrees of freedom from weak to strong coupling.

For close-to-conformal evolution, this dependence on the number of degrees of freedom can be factored out in such a way that the final spectrum becomes a shift of the static thermal spectrum. By comparing the weak and strong coupling calculations of the rescaled spectrum or, equivalently, the fixed temperature spectrum, we find that the weak coupling calculation $\lambda=0.5$ result reproduces even quantitatively the strong coupling result. This is an interesting observation, given that $\lambda=0.5$ marks a limit of applicability of the perturbative computation, since for this value the Debye mass equals the temperature. While this may just be a numerical coincidence, it is worth mentioning that at values of the coupling as large as $\lambda=0.5$ the matter fields of the theory acquire a large thermal width, which obscures their interpretation as quasi-particle excitations. It would be interesting to test whether the strong coupling limit may be mimicked by leading-order results at some fixed value of the coupling constant for other gauge theories in which the strong coupling limit can be addressed via the gauge/gravity duality, like for example~\cite{Polchinski:2000uf,Pilch:2000ue,Klebanov:2000hb,Maldacena:2000yy}.\\

\noindent
{\bf Acknowledgements.}~We thank D.~Mateos and B.~Fiol for useful discussions. We  are supported by grants SGR-2017-754, PID2019-105614GB-C21  and the ``Unit of Excellence MdM 2020-2023'' award to the Institute of Cosmos Sciences (CEX2019-000918-M).

\bibliographystyle{jhep}
\bibliography{biblio/biblio.bib}

\providecommand{\href}[2]{#2}\begingroup\raggedright\begin{thebibliography}{10}

\bibitem{LIGOScientific:2016aoc}
{\scshape LIGO Scientific, Virgo} collaboration, B.~P. Abbott et~al.,
  \emph{{Observation of Gravitational Waves from a Binary Black Hole Merger}},
  \href{http://dx.doi.org/10.1103/PhysRevLett.116.061102}{\emph{Phys. Rev.
  Lett.} {\bf 116} (2016) 061102}, [\href{http://arxiv.org/abs/1602.03837}{{\tt
  1602.03837}}].

\bibitem{LIGOScientific:2020tif}
{\scshape LIGO Scientific, Virgo} collaboration, R.~Abbott et~al., \emph{{Tests
  of general relativity with binary black holes from the second LIGO-Virgo
  gravitational-wave transient catalog}},
  \href{http://dx.doi.org/10.1103/PhysRevD.103.122002}{\emph{Phys. Rev. D} {\bf
  103} (2021) 122002}, [\href{http://arxiv.org/abs/2010.14529}{{\tt
  2010.14529}}].

\bibitem{Barack:2018yly}
L.~Barack et~al., \emph{{Black holes, gravitational waves and fundamental
  physics: a roadmap}},
  \href{http://dx.doi.org/10.1088/1361-6382/ab0587}{\emph{Class. Quant. Grav.}
  {\bf 36} (2019) 143001}, [\href{http://arxiv.org/abs/1806.05195}{{\tt
  1806.05195}}].

\bibitem{Aggarwal:2020olq}
N.~Aggarwal et~al., \emph{{Challenges and opportunities of gravitational-wave
  searches at MHz to GHz frequencies}},
  \href{http://dx.doi.org/10.1007/s41114-021-00032-5}{\emph{Living Rev. Rel.}
  {\bf 24} (2021) 4}, [\href{http://arxiv.org/abs/2011.12414}{{\tt
  2011.12414}}].

\bibitem{Caprini:2018mtu}
C.~Caprini and D.~G. Figueroa, \emph{{Cosmological Backgrounds of Gravitational
  Waves}}, \href{http://dx.doi.org/10.1088/1361-6382/aac608}{\emph{Class.
  Quant. Grav.} {\bf 35} (2018) 163001},
  [\href{http://arxiv.org/abs/1801.04268}{{\tt 1801.04268}}].

\bibitem{weinberg1972gravitation}
S.~Weinberg, \emph{Gravitation and cosmology: principles and applications of
  the general theory of relativity}, .

\bibitem{ghiglieri2015gravitational}
J.~Ghiglieri and M.~Laine, \emph{{Gravitational wave background from Standard
  Model physics: Qualitative features}},
  \href{http://dx.doi.org/10.1088/1475-7516/2015/07/022}{\emph{JCAP} {\bf 07}
  (2015) 022}, [\href{http://arxiv.org/abs/1504.02569}{{\tt 1504.02569}}].

\bibitem{ghiglieri2020gravitational}
J.~Ghiglieri, G.~Jackson, M.~Laine and Y.~Zhu, \emph{{Gravitational wave
  background from Standard Model physics: Complete leading order}},
  \href{http://dx.doi.org/10.1007/JHEP07(2020)092}{\emph{JHEP} {\bf 07} (2020)
  092}, [\href{http://arxiv.org/abs/2004.11392}{{\tt 2004.11392}}].

\bibitem{ringwald2021gravitational}
A.~Ringwald, J.~Sch\"utte-Engel and C.~Tamarit, \emph{{Gravitational Waves as a
  Big Bang Thermometer}},
  \href{http://dx.doi.org/10.1088/1475-7516/2021/03/054}{\emph{JCAP} {\bf 03}
  (2021) 054}, [\href{http://arxiv.org/abs/2011.04731}{{\tt 2011.04731}}].

\bibitem{Ghiglieri:2020dpq}
J.~Ghiglieri, A.~Kurkela, M.~Strickland and A.~Vuorinen, \emph{{Perturbative
  Thermal QCD: Formalism and Applications}},
  \href{http://dx.doi.org/10.1016/j.physrep.2020.07.004}{\emph{Phys. Rept.}
  {\bf 880} (2020) 1--73}, [\href{http://arxiv.org/abs/2002.10188}{{\tt
  2002.10188}}].

\bibitem{Ghiglieri:2013gia}
J.~Ghiglieri, J.~Hong, A.~Kurkela, E.~Lu, G.~D. Moore and D.~Teaney,
  \emph{{Next-to-leading order thermal photon production in a weakly coupled
  quark-gluon plasma}},
  \href{http://dx.doi.org/10.1007/JHEP05(2013)010}{\emph{JHEP} {\bf 05} (2013)
  010}, [\href{http://arxiv.org/abs/1302.5970}{{\tt 1302.5970}}].

\bibitem{Cacciapaglia:2020kgq}
G.~Cacciapaglia, C.~Pica and F.~Sannino, \emph{{Fundamental Composite Dynamics:
  A Review}},
  \href{http://dx.doi.org/10.1016/j.physrep.2020.07.002}{\emph{Phys. Rept.}
  {\bf 877} (2020) 1--70}, [\href{http://arxiv.org/abs/2002.04914}{{\tt
  2002.04914}}].

\bibitem{Maldacena:1997re}
J.~M. Maldacena, \emph{{The Large N limit of superconformal field theories and
  supergravity}}, \href{http://dx.doi.org/10.1023/A:1026654312961}{\emph{Adv.
  Theor. Math. Phys.} {\bf 2} (1998) 231--252},
  [\href{http://arxiv.org/abs/hep-th/9711200}{{\tt hep-th/9711200}}].

\bibitem{Aharony:1999ti}
O.~Aharony, S.~S. Gubser, J.~M. Maldacena, H.~Ooguri and Y.~Oz, \emph{{Large N
  field theories, string theory and gravity}},
  \href{http://dx.doi.org/10.1016/S0370-1573(99)00083-6}{\emph{Phys. Rept.}
  {\bf 323} (2000) 183--386}, [\href{http://arxiv.org/abs/hep-th/9905111}{{\tt
  hep-th/9905111}}].

\bibitem{Witten:1998qj}
E.~Witten, \emph{{Anti-de Sitter space and holography}},
  \href{http://dx.doi.org/10.4310/ATMP.1998.v2.n2.a2}{\emph{Adv. Theor. Math.
  Phys.} {\bf 2} (1998) 253--291},
  [\href{http://arxiv.org/abs/hep-th/9802150}{{\tt hep-th/9802150}}].

\bibitem{Gubser:1998bc}
S.~S. Gubser, I.~R. Klebanov and A.~M. Polyakov, \emph{{Gauge theory
  correlators from noncritical string theory}},
  \href{http://dx.doi.org/10.1016/S0370-2693(98)00377-3}{\emph{Phys. Lett. B}
  {\bf 428} (1998) 105--114}, [\href{http://arxiv.org/abs/hep-th/9802109}{{\tt
  hep-th/9802109}}].

\bibitem{Son:2002sd}
D.~T. Son and A.~O. Starinets, \emph{{Minkowski space correlators in AdS / CFT
  correspondence: Recipe and applications}},
  \href{http://dx.doi.org/10.1088/1126-6708/2002/09/042}{\emph{JHEP} {\bf 09}
  (2002) 042}, [\href{http://arxiv.org/abs/hep-th/0205051}{{\tt
  hep-th/0205051}}].

\bibitem{kovtun2005quasinormal}
P.~K. Kovtun and A.~O. Starinets, \emph{{Quasinormal modes and holography}},
  \href{http://dx.doi.org/10.1103/PhysRevD.72.086009}{\emph{Phys. Rev. D} {\bf
  72} (2005) 086009}, [\href{http://arxiv.org/abs/hep-th/0506184}{{\tt
  hep-th/0506184}}].

\bibitem{laine2016basics}
M.~Laine and A.~Vuorinen, \emph{{Basics of Thermal Field Theory}},
  \href{http://dx.doi.org/10.1007/978-3-319-31933-9}{\emph{Lect. Notes Phys}
  {\bf 925} (2016) 1--281}, [\href{http://arxiv.org/abs/1701.01554}{{\tt
  1701.01554}}].

\bibitem{Teaney:2006nc}
D.~Teaney, \emph{{Finite temperature spectral densities of momentum and
  R-charge correlators in N=4 Yang Mills theory}},
  \href{http://dx.doi.org/10.1103/PhysRevD.74.045025}{\emph{Phys. Rev. D} {\bf
  74} (2006) 045025}, [\href{http://arxiv.org/abs/hep-ph/0602044}{{\tt
  hep-ph/0602044}}].

\bibitem{Kovtun:2006pf}
P.~Kovtun and A.~Starinets, \emph{{Thermal spectral functions of strongly
  coupled N=4 supersymmetric Yang-Mills theory}},
  \href{http://dx.doi.org/10.1103/PhysRevLett.96.131601}{\emph{Phys. Rev.
  Lett.} {\bf 96} (2006) 131601},
  [\href{http://arxiv.org/abs/hep-th/0602059}{{\tt hep-th/0602059}}].

\bibitem{morse1954methods}
P.~M. Morse and H.~Feshbach, \emph{Methods of theoretical physics},
  {\emph{American Journal of Physics} {\bf 22} (1954) 410--413}.

\bibitem{Callan:1970ze}
C.~G. Callan, Jr., S.~R. Coleman and R.~Jackiw, \emph{{A New improved energy -
  momentum tensor}},
  \href{http://dx.doi.org/10.1016/0003-4916(70)90394-5}{\emph{Annals Phys.}
  {\bf 59} (1970) 42--73}.

\bibitem{Fiol:2019woe}
B.~Fiol and J.~Mart\'\i{}nez-Montoya, \emph{{On scalar radiation}},
  \href{http://dx.doi.org/10.1007/JHEP03(2020)087}{\emph{JHEP} {\bf 03} (2020)
  087}, [\href{http://arxiv.org/abs/1907.08161}{{\tt 1907.08161}}].

\bibitem{Yamada:2006rx}
D.~Yamada and L.~G. Yaffe, \emph{{Phase diagram of N=4 super-Yang-Mills theory
  with R-symmetry chemical potentials}},
  \href{http://dx.doi.org/10.1088/1126-6708/2006/09/027}{\emph{JHEP} {\bf 09}
  (2006) 027}, [\href{http://arxiv.org/abs/hep-th/0602074}{{\tt
  hep-th/0602074}}].

\bibitem{du2020two}
Q.~Du, M.~Strickland, U.~Tantary and B.-W. Zhang, \emph{{Two-loop HTL-resummed
  thermodynamics for $ \mathcal{N} $ = 4 supersymmetric Yang-Mills theory}},
  \href{http://dx.doi.org/10.1007/JHEP09(2020)038}{\emph{JHEP} {\bf 09} (2020)
  038}, [\href{http://arxiv.org/abs/2006.02617}{{\tt 2006.02617}}].

\bibitem{Gubser:1996de}
S.~S. Gubser, I.~R. Klebanov and A.~W. Peet, \emph{{Entropy and temperature of
  black 3-branes}},
  \href{http://dx.doi.org/10.1103/PhysRevD.54.3915}{\emph{Phys. Rev. D} {\bf
  54} (1996) 3915--3919}, [\href{http://arxiv.org/abs/hep-th/9602135}{{\tt
  hep-th/9602135}}].

\bibitem{Saikawa:2018rcs}
K.~Saikawa and S.~Shirai, \emph{{Primordial gravitational waves, precisely: The
  role of thermodynamics in the Standard Model}},
  \href{http://dx.doi.org/10.1088/1475-7516/2018/05/035}{\emph{JCAP} {\bf 05}
  (2018) 035}, [\href{http://arxiv.org/abs/1803.01038}{{\tt 1803.01038}}].

\bibitem{casalderrey2014gauge}
J.~Casalderrey-Solana, H.~Liu, D.~Mateos, K.~Rajagopal and U.~A. Wiedemann,
  \emph{{Gauge/String Duality, Hot QCD and Heavy Ion Collisions}},
  \href{http://dx.doi.org/10.1017/CBO9781139136747}{\emph{UK: Cambridge
  University Press} (2014) }, [\href{http://arxiv.org/abs/1101.0618}{{\tt
  1101.0618}}].

\bibitem{Caron-Huot:2006pee}
S.~Caron-Huot, P.~Kovtun, G.~D. Moore, A.~Starinets and L.~G. Yaffe,
  \emph{{Photon and dilepton production in supersymmetric Yang-Mills plasma}},
  \href{http://dx.doi.org/10.1088/1126-6708/2006/12/015}{\emph{JHEP} {\bf 12}
  (2006) 015}, [\href{http://arxiv.org/abs/hep-th/0607237}{{\tt
  hep-th/0607237}}].

\bibitem{Bea:2021zol}
Y.~Bea, J.~Casalderrey-Solana, T.~Giannakopoulos, A.~Jansen, S.~Krippendorf,
  D.~Mateos et~al., \emph{{Spinodal Gravitational Waves}},
  \href{http://arxiv.org/abs/2112.15478}{{\tt 2112.15478}}.

\bibitem{Bea:2021zsu}
Y.~Bea, J.~Casalderrey-Solana, T.~Giannakopoulos, D.~Mateos,
  M.~Sanchez-Garitaonandia and M.~Zilh\~ao, \emph{{Bubble wall velocity from
  holography}},
  \href{http://dx.doi.org/10.1103/PhysRevD.104.L121903}{\emph{Phys. Rev. D}
  {\bf 104} (2021) L121903}, [\href{http://arxiv.org/abs/2104.05708}{{\tt
  2104.05708}}].

\bibitem{Ares:2020lbt}
F.~R. Ares, M.~Hindmarsh, C.~Hoyos and N.~Jokela, \emph{{Gravitational waves
  from a holographic phase transition}},
  \href{http://dx.doi.org/10.1007/JHEP04(2021)100}{\emph{JHEP} {\bf 21} (2020)
  100}, [\href{http://arxiv.org/abs/2011.12878}{{\tt 2011.12878}}].

\bibitem{Ares:2021ntv}
F.~R. Ares, O.~Henriksson, M.~Hindmarsh, C.~Hoyos and N.~Jokela,
  \emph{{Effective actions and bubble nucleation from holography}},
  \href{http://arxiv.org/abs/2109.13784}{{\tt 2109.13784}}.

\bibitem{Ares:2021nap}
F.~R. Ares, O.~Henriksson, M.~Hindmarsh, C.~Hoyos and N.~Jokela,
  \emph{{Gravitational Waves at Strong Coupling from an Effective Action}},
  \href{http://arxiv.org/abs/2110.14442}{{\tt 2110.14442}}.

\bibitem{Bigazzi:2020avc}
F.~Bigazzi, A.~Caddeo, A.~L. Cotrone and A.~Paredes, \emph{{Dark Holograms and
  Gravitational Waves}},
  \href{http://dx.doi.org/10.1007/JHEP04(2021)094}{\emph{JHEP} {\bf 04} (2021)
  094}, [\href{http://arxiv.org/abs/2011.08757}{{\tt 2011.08757}}].

\bibitem{Bigazzi:2021ucw}
F.~Bigazzi, A.~Caddeo, T.~Canneti and A.~L. Cotrone, \emph{{Bubble wall
  velocity at strong coupling}},
  \href{http://dx.doi.org/10.1007/JHEP08(2021)090}{\emph{JHEP} {\bf 08} (2021)
  090}, [\href{http://arxiv.org/abs/2104.12817}{{\tt 2104.12817}}].

\bibitem{Arnold:2000dr}
P.~B. Arnold, G.~D. Moore and L.~G. Yaffe, \emph{{Transport coefficients in
  high temperature gauge theories. 1. Leading log results}},
  \href{http://dx.doi.org/10.1088/1126-6708/2000/11/001}{\emph{JHEP} {\bf 11}
  (2000) 001}, [\href{http://arxiv.org/abs/hep-ph/0010177}{{\tt
  hep-ph/0010177}}].

\bibitem{Polchinski:2000uf}
J.~Polchinski and M.~J. Strassler, \emph{{The String dual of a confining
  four-dimensional gauge theory}},
  \href{http://arxiv.org/abs/hep-th/0003136}{{\tt hep-th/0003136}}.

\bibitem{Pilch:2000ue}
K.~Pilch and N.~P. Warner, \emph{{N=2 supersymmetric RG flows and the IIB
  dilaton}}, \href{http://dx.doi.org/10.1016/S0550-3213(00)00656-8}{\emph{Nucl.
  Phys. B} {\bf 594} (2001) 209--228},
  [\href{http://arxiv.org/abs/hep-th/0004063}{{\tt hep-th/0004063}}].

\bibitem{Klebanov:2000hb}
I.~R. Klebanov and M.~J. Strassler, \emph{{Supergravity and a confining gauge
  theory: Duality cascades and chi SB resolution of naked singularities}},
  \href{http://dx.doi.org/10.1088/1126-6708/2000/08/052}{\emph{JHEP} {\bf 08}
  (2000) 052}, [\href{http://arxiv.org/abs/hep-th/0007191}{{\tt
  hep-th/0007191}}].

\bibitem{Maldacena:2000yy}
J.~M. Maldacena and C.~Nunez, \emph{{Towards the large N limit of pure N=1
  superYang-Mills}},
  \href{http://dx.doi.org/10.1103/PhysRevLett.86.588}{\emph{Phys. Rev. Lett.}
  {\bf 86} (2001) 588--591}, [\href{http://arxiv.org/abs/hep-th/0008001}{{\tt
  hep-th/0008001}}].

\end{thebibliography}\endgroup


\end{document}